\documentclass{article}
\usepackage{arxiv}

\usepackage{amssymb}
\usepackage{authblk}                %
\usepackage[numbers,sort&compress]{natbib}                %
\usepackage{hyperref}               %
\usepackage{amsmath}
\usepackage{amssymb}
\usepackage{xcolor}
\usepackage{verbatim}
\usepackage{algorithmic}
\usepackage{url}
\usepackage{graphicx}
\usepackage{textcomp}
\usepackage{multirow}
\usepackage{pdfpages}
\usepackage{xcolor}
\usepackage{tikz}
\usepackage{setspace}
\usepackage{float}
\usepackage{booktabs}
\usepackage{tabularx}
\usepackage{booktabs}
\usepackage{pdfpages}
\usepackage{rotating}

\title{Analysis of Architecture Options for Foundation-model-based Agents: A Taxonomy and Decision Model}
 
\author[1,2]{Jingwen Zhou}
\author[2,3]{Qinghua Lu}
\author[2]{Jieshan Chen}
\author[2,3]{Liming Zhu}
\author[2]{Xiwei Xu}
\author[2,4]{Zhenchang Xing}
\author[2]{Stefan Harrer}
 
\affil[1]{School of Computer, Data and Mathematical Sciences, Western Sydney University, Sydney, Australia}
\affil[2]{CSIRO's Data61, Australia}
\affil[3]{University of New South Wales, Sydney, New South Wales, Australia}
\affil[4]{Australian National University, Canberra, Australian Capital Territory, Australia}
 
\date{}
 
\begin{document}
 
\maketitle
 
\renewcommand{\thefootnote}{}
\footnotetext{\textit{Email addresses:} \texttt{helen.zhou@data61.csiro.au} (Jingwen Zhou),
\texttt{qinghua.lu@data61.csiro.au} (Qinghua Lu),
\texttt{jieshan.chen@data61.csiro.au} (Jieshan Chen),
\texttt{liming.zhu@data61.csiro.au} (Liming Zhu),
\texttt{xiwei.xu@data61.csiro.au} (Xiwei Xu),
\texttt{zhenchang.xing@data61.csiro.au} (Zhenchang Xing),
\texttt{stefan.harrer@data61.csiro.au} (Stefan Harrer)}
\renewcommand{\thefootnote}{\arabic{footnote}}
 
\begin{abstract}
The rapid advancement of foundation models has led to a new generation of AI-based agents. \textcolor{black}{While mature frameworks exist, the system-level architectural decision-making is still often ad-hoc,} as architects lack a systematic, evidence-based framework for navigating the complex landscape of architectural options and their resulting trade-offs. To address this critical gap, we conducted a Systematic Literature Review (SLR) of 66 primary studies to develop two key artifacts: (1) a comprehensive taxonomy that systematically classifies architectural design options for foundation-model-based agents across functional capabilities and non-functional qualities, and (2) an accompanying decision model, presented as a matrix, that maps these options to their impact on system qualities, making critical design trade-offs explicit. The taxonomy and decision model were then validated through a multi-faceted process, including a formal survey with 12 experts from both academia and industry. This work provides architects with a common vocabulary and a structured guide for designing and reasoning about agent systems, thereby providing a rigorous foundation to address the current fragmentation in the field.
\end{abstract}
 
 
\bigskip

\section{Introduction}
\label{sec:intro}
The emergence of Foundation Models (FMs)—large-scale models pre-trained on vast amounts of data represents a fundamental paradigm shift in Artificial Intelligence~\cite{lu2024towards}. Their ability to adapt to a wide range of downstream tasks has enabled a new class of sophisticated applications, chief among them being Foundation-Model-based agents (FM-based agents). These are autonomous systems that leverage FMs as their core cognitive engine to perceive, reason, plan, and act to achieve complex goals, with the potential to revolutionize industries from healthcare to smart manufacturing~\cite{lu2024towards, budler2023review, masterman2024landscape}. The rapid proliferation of these agents, exemplified by systems like Auto-GPT\footnote{https://github.com/Significant-Gravitas/AutoGPT} and BabyAGI\footnote{https://github.com/yoheinakajima/babyagi} for autonomous task management, Voyager for embodied learning~\cite{wang2023voyager}, UI-based computational agents like the OpenAI Operator\footnote{https://openai.com/index/introducing-operator/}, \textcolor{black}{Claude Opus 5\footnote{https://www.anthropic.com/news/claude-opus-5} that interact with computer systems through natural language interfaces}, the first AI software engineer Devin\footnote{https://devin.ai/}, and open-source local execution tools like Open Interpreter\footnote{https://www.openinterpreter.com/}, signals a move towards more capable and independent AI systems.

\subsection{\textcolor{black}{Motivation}}
However, this increasing capability brings significant architectural challenges. Today's agents employ cutting-edge technologies” and “new operational paradigms” that stretch the boundaries of traditional software design~\cite{lu2024towards}. For instance, multi-agent frameworks like AutoGen~\cite{wu2023autogen} introduce complex architectural challenges in coordination of conversational workflows, ensuring effective communication protocols between agents, and maintaining a consistent state as multiple agents collaboratively attempt to complete a task. At the same time, agents like Voyager possess novel capabilities such as long-term \textit{memory} and \textit{reflection}, creating new design problems related to efficiency, and observability~\cite{wang2023voyager}. \textcolor{black}{While structured frameworks like LangChain and Microsoft Semantic Kernel provide strong operational foundations, the higher-level architectural decision-making process remains fragmented. Consequently, system-level design choices are often ad-hoc, leaving architects without a common vocabulary and a systematic framework to reason about the trade-offs of their choices.}


To fully appreciate these challenges, it is crucial to clarify what constitutes the \textbf{“architectures and frameworks”} in this context. An agent's \textbf{architecture} refers to the fundamental components that constitute the system (e.g., planning, memory, action modules) and their relationships. The \textbf{frameworks} that have emerged to build these systems, such as the aforementioned AutoGen~\cite{wu2023autogen}, LangChain\footnote{https://www.langchain.com/}, data-centric solutions like LlamaIndex\footnote{https://www.llamaindex.ai/}, orchestration frameworks like CrewAI\footnote{https://www.crewai.com/}, autonomous systems like SuperAGI\footnote{https://superagi.com/}, and platform-integrated solutions like the Microsoft Copilot Framework\footnote{https://news.microsoft.com/source/features/ai/microsoft-outlines-framework-for-building-ai-apps-and-copilots-expands-ai-plugin-ecosystem/}, provide reusable components and abstractions that expose this vast and complex design space to developers.

The proliferation of these powerful frameworks and the novel architectural patterns they enable has led to a fragmented landscape of ad-hoc designs. Architects lack a common vocabulary and a systematic method to reason about the trade-offs of their choices. This gap in shared architectural knowledge directly hinders the \textbf{“optimization of an architecture for improved functionality,”} which is the process of making conscious design choices to better achieve desired system behaviors (like autonomous learning) and quality attributes (like reliability). Without a clear understanding of the trade-offs for different architectural options, it is difficult to analyze and ensure these critical system-level qualities. A comprehensive taxonomy is therefore needed to provide the structured, evidence-based foundation for designing and analyzing these complex systems.

\subsection{\textcolor{black}{Scope and Contributions}}
\textcolor{black}{To address this architectural gap, we introduce a comprehensive, evidence-based taxonomy of 74 architectural design options for FM-based agents, derived from a systematic literature review (SLR) of 66 primary studies. In this paper, we define an \textbf{architecture option} as a specific, fundamental design approach or pattern adopted to address a particular architectural concern (e.g., selecting a 'Centralized' vs. 'Fully Distributed' coordination mechanism). It is a more granular and concrete choice than a general "design decision"~\cite{jansen2005software}, as it refers to a selection from a predefined set of architectural alternatives within our taxonomy. Paired with this taxonomy, we introduce a decision model to analyze the trade-offs between these options, offering deeper insights and guiding the design of the next generation of FM-based agents. The main contributions of this paper are as follows:}

\begin{itemize}
    \item \textcolor{black}{\textbf{A Comprehensive Taxonomy of Architectural Options:} We present a detailed, evidence-based taxonomy for FM-based agents, derived from a systematic literature review and refined through expert validation. The taxonomy systematically classifies architectural options across functional capabilities and non-functional qualities.}
    \item \textcolor{black}{\textbf{An Architectural Decision Model:} We introduce a practical decision model, presented as a Decision Matrix, that analyzes the trade-offs between architectural choices and their impact on key system qualities to guide architects in their design process.}
    \item \textbf{Validation and discussion of the taxonomy's implications:} The taxonomy and decision model underwent a multi-faceted validation process, including a formal survey with 12 experts from both academia and industry. Furthermore, we articulate the taxonomy and decision model's specific implications for key stakeholders, including researchers, architects, and tool builders, clarifying its value for both academic research and industrial practice.
\end{itemize}

\textcolor{black}{Specifically, Section~\ref{sec:related} reviews related work. Section~\ref{sec:methodology} details our systematic methodology. The complete taxonomy is presented in Section~\ref{sec:taxonomy}, followed by the decision model in Section~\ref{sec:decision_model}. We describe our validation of these artifacts in Section~\ref{sec:validation}. We discuss threats to validity in Section~\ref{sec:threats}, and introduce the implications for key stakeholders in Section~\ref{sec:discussion}. Finally, Section \ref{sec:con} concludes the paper, summarizing our key findings and outlining future research directions.}

\section{Background and Related Work}
\label{sec:related}
\subsection{FM-based Agent Frameworks}
The landscape of FM-based agents is characterized by rapid advancements in both the foundational models that power them and the frameworks designed to orchestrate their behavior. Major technology companies have developed increasingly powerful Large Language Models (LLMs) that serve as the core cognitive engines for modern agents. \textcolor{black}{For instance, Google's Gemini 3.6 Flash is optimised for reasoning, coding, and agentic tasks, and has been integrated across their products \cite{gemini2026}. Similarly, Meta's Llama 4 represents another highly efficient and scalable foundation model, adopting a mixture-of-experts architecture to support long-context and multimodal tasks \cite{llama4}.} 

\textcolor{black}{Building upon these powerful models, a rich ecosystem of agent frameworks has emerged to tackle complex, multi-step tasks. A significant focus has been on multi-agent collaboration. Microsoft's \textbf{AutoGen} is a prominent open-source framework designed to facilitate complex conversations and cooperation between multiple AI agents \cite{wu2023autogen, taskade2024top}. \textbf{MetaGPT} takes a novel approach by emulating a software company's structure, assigning roles like project managers and engineers to different agents to collaboratively develop code \cite{hong2023metagpt}. \textbf{CrewAI} also focuses on enhancing multi-agent collaboration, enabling agents to work together to solve tasks \cite{crewai2023}. Distinct from these, frameworks like \textbf{LangChain} provide a general-purpose toolkit for developers, focusing on integrating LLMs with a wide array of external tools and data sources to enable the seamless execution of complex workflows.}

\subsection{Existing Surveys and Positioning of Our Work}
\label{subsec:positioning}
\textcolor{black}{Despite the proliferation of powerful agent frameworks, a notable gap persists in the systematic analysis and standardization of their underlying architectures. While specific platforms like AIOS demonstrate how to create cohesive environments for autonomous agents \cite{mei2024aios}, and frameworks such as MAToM-SNN leverage novel neural networks for complex agent interactions \cite{zhao2023brain}, they represent bespoke implementations rather than generalizable architectural knowledge that can guide future designs. Several recent surveys have highlighted this challenge from different perspectives. For instance, some reviews provide a broad overview of agent frameworks and their capabilities, confirming the lack of standardized architectures as a barrier to integration \cite{wang2024survey}. Other studies focus on specific aspects like prompting techniques \cite{schulhoff2024prompt} or the challenges in multi-agent systems \cite{guo2024large}, often concluding with a call for more robust and standardized frameworks \cite{xi2023rise, li2024survey}.}

\textcolor{black}{While existing studies correctly identify the need for standardization, our work addresses a more specific and fundamental gap: the lack of a comprehensive, evidence-based taxonomy from a \textbf{software architecture perspective}. To our knowledge, no prior work has systematically catalogued the fundamental \textit{architectural design options} for FM-based agents and explicitly analyzed their resulting \textit{non-functional trade-offs}. Our work aims to fill this gap by offering a detailed taxonomy and a corresponding decision model. The framework provides a systematic approach by mapping these architectural options to a comprehensive set of non-functional qualities, ranging from core engineering attributes like Performance, Reliability, Robustness, Resilience, Security, Maintainability, and Compatibility, to human-centric concerns such as Usability and Responsible \& Trustworthy AI.}
\section{Methodology: A Systematic Process for Taxonomy Development}
\label{sec:methodology}
To construct a rigorous and evidence-based taxonomy for FM-based agent architectures, we adopted the systematic taxonomy development method proposed by Usman et al.~\cite{usman2017taxonomies}. This established process provides a transparent and reproducible framework, ensuring that our taxonomy is grounded in a comprehensive analysis of the existing literature. The methodology consists of four main phases---Planning, Identification \& Extraction, Design \& Construction, and Validation---which are detailed in the following subsections. The initial evidence collection for this process was conducted through the SLR~\cite{ZHANG2011625}, which allowed us to systematically gather and analyze architectural characteristics from primary studies. This entire process is guided by our research questions:

\begin{itemize}
    \item \textcolor{black}{\textbf{RQ1: What are the fundamental architectural elements and design options that characterize existing FM-based agents?}}
    
    \textcolor{black}{This initial question drives the descriptive phase of our systematic literature review. It aims to identify and catalogue not just the capabilities of agents, but the core building blocks—the foundational components, patterns, and configurable choices—that constitute their \textit{architecture}.}
    
    \item \textcolor{black}{\textbf{RQ2: How do the choices among these architectural elements create design trade-offs and influence system-level qualities?}}
    
    \textcolor{black}{This question moves from description to analysis, targeting the core of architectural reasoning. It investigates the consequences of design decisions, seeking to understand how selecting a specific architectural option (e.g., a centralized coordinator) impacts critical system-level qualities (e.g., performance, scalability, reliability) and creates inherent \textit{architectural trade-offs}.}
    
    \item \textcolor{black}{\textbf{RQ3: How can a structured classification of these elements and trade-offs be utilized to guide the architectural design and decision-making for FM-based agents?}}
    
    \textcolor{black}{This final question addresses the practical utility of our work. It assesses whether a systematic taxonomy of the identified elements (from RQ1) and their associated trade-offs (from RQ2) can serve as an effective framework to not only analyze existing systems but also to \textit{guide architects} in designing the next generation of FM-based agents, a question directly addressed by our proposed Decision Model.}
\end{itemize}

\subsection{Planning}

\textcolor{black}{The planning phase defines the taxonomy's context and initial setting. According to the guidelines from Usman et al.~\cite{usman2017taxonomies}, this phase consists of five key activities that establish the foundation for the entire development process.}

\begin{figure*}
    \centering
    \includegraphics[width=0.95\linewidth]{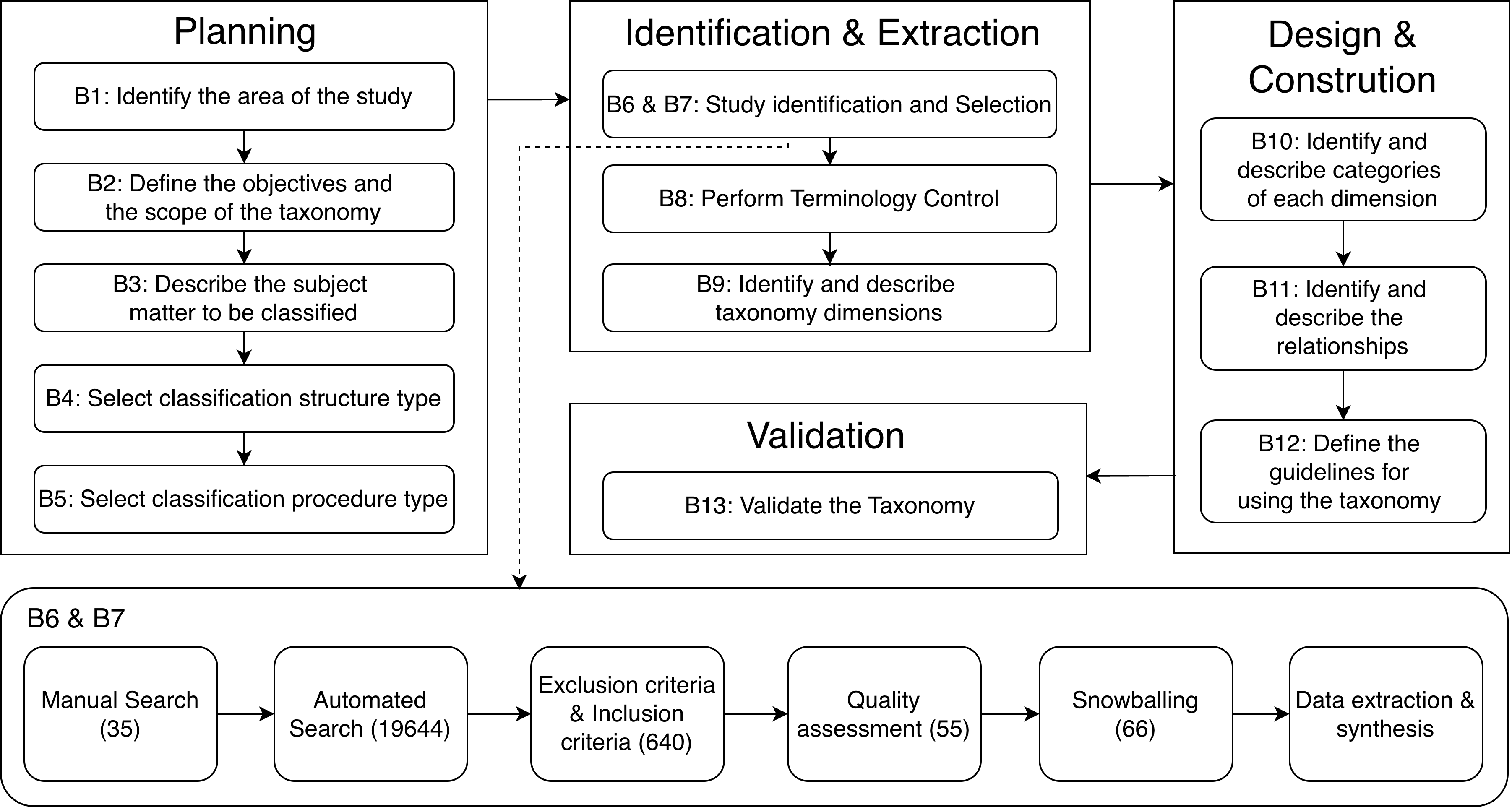}
    \caption{\textcolor{black}{Methodology. The steps labeled B1-B13 are adopted from the taxonomy development process proposed by Usman et al.~\cite{usman2017taxonomies}.}}
    \label{fig:methodology}
\end{figure*}

\subsubsection{B1: Identify the Area of the Study}
\textcolor{black}{The knowledge area for this taxonomy is Software Engineering, with a specific focus on the software architecture of FM-based agents.}

\subsubsection{B2: Define the Objectives and the Scope of the Taxonomy}
\textcolor{black}{The primary objective of this study is to develop a comprehensive taxonomy that systematically classifies the architectural options for FM-based agents. The taxonomy is designed to address challenges in the design and implementation of these agents by integrating both functional and non-functional aspects. Its scope is therefore defined to include architectural decisions at both \textbf{design-time} (e.g., selecting a coordination model) and \textbf{run-time} (e.g., dynamic tool selection), covering the key stages of an agent's lifecycle.}

\subsubsection{B3: Describe the Subject Matter to be Classified}
\textcolor{black}{The subject matter of this taxonomy is the architecture options for FM-based agents. As defined in Section 1.2, an architecture option is a specific design approach adopted to address a particular architectural concern. Our subsequent SLR aims to identify, extract, and synthesize concrete instances of these architectural options from the existing literature, encompassing both the core operational mechanisms and the system-level quality attributes that characterize them.}

\subsubsection{B4: Select Classification Structure Type}

\textcolor{black}{This study adopts a \textbf{faceted taxonomy} approach \cite{kwasnik1999role, prieto1991implementing}. This structure supports classification across multiple independent dimensions and is well-suited to the evolving nature of FM-based agents, as it provides flexibility in representing their diverse attributes. The two primary facets of our taxonomy—\textbf{Functional Capabilities} and \textbf{Non-functional Qualities}—directly correspond to the dimensions of the subject matter identified in B3.}

\subsubsection{B5: Select Classification Procedure Type}
This study employs a qualitative classification procedure, utilizing \textbf{thematic analysis} as the core technique to synthesize the data extracted from our SLR \cite{glass1995contemporary, wheaton1968development}. Specifically, in the Design \& Construction phase (Section \ref{subsec:design-construction}), we apply an iterative process of coding and grouping to the extracted architectural features. This process allows us to systematically identify recurring themes, which then form the basis for defining the dimensions (Step B9) and categories (Step B10) of our taxonomy. 
\textcolor{black}{To mitigate individual bias, we adopted a negotiated agreement approach for the qualitative classification.} The initial classification was performed by one researcher and then thoroughly reviewed by a second researcher. Any disagreements were subsequently resolved through discussion and consensus among all authors.

\subsection{Identification \& Extraction}
\subsubsection{B6 \& B7: Study Identification and Selection}
\textcolor{black}{To ensure a comprehensive and replicable selection of primary studies, we adopted the Quasi-Gold Standard (QGS) search strategy approach~\cite{ZHANG2011625}. This is a well-established methodology in software engineering systematic reviews that aims to maximize coverage and relevance. As illustrated in the exemplar study by Hou et al.~\cite{hou2024large}, the QGS process involves three main steps:} 

\begin{itemize}
    \item \textcolor{black}{A manual search of top-tier venues to build a "quasi-gold standard" set of highly relevant papers.}
    \item \textcolor{black}{Derivation of a comprehensive search string based on the keywords and terminology used in the QGS set.}
    \item \textcolor{black}{Execution of a broad automated search across multiple digital libraries using the derived search string.}
\end{itemize}

\paragraph{\textbf{Manual Search:}} We began by performing a manual search in top-tier software engineering venues, selected for their high impact and relevance to software architecture and engineering research. The selected venues are: ASE (IEEE/ACM International Conference on Automated Software Engineering), FSE (ACM SIGSOFT Symposium on the Foundations of Software Engineering), TOSEM (ACM Transactions on Software Engineering and Methodology), ACM Computing Surveys, ICSE (IEEE/ACM International Conference on Software Engineering), ISSTA (ACM SIGSOFT International Symposium on Software Testing and Analysis), TSE (IEEE Transactions on Software Engineering), and JSS (Journal of Systems and Software). From these venues, we systematically identified an initial set of 35 primary studies that directly addressed the architecture of FM-based agents. This collection of 35 papers served as our Quasi-Gold Standard (QGS) set, forming an evidence-based foundation for developing our search string. Using the 35 papers in the QGS set as a ground truth, \textcolor{black}{we performed a systematic keyword analysis with the assistance of a Large Language Model (LLM) to systematically extract and categorize prevalent terms and their synonyms (e.g., specific SE tasks like automated program repair) from the QGS texts. The detailed few-shot extraction prompt and the complete data logs are publicly available in our replication package.} The final search string is composed of two main conceptual pillars, designed to be combined with a Boolean AND operator. Pillar A covers the core technology (the agent), while Pillar B covers the domain and discipline (the architecture and SE tasks). The keywords within each pillar are combined with a Boolean OR operator.

\begin{itemize}
    \item \textbf{Pillar A: The Technology (FM-based Agent)} \\
    \textcolor{black}{Keywords included: \textit{Foundation Models, Large Language Models (LLMs), Generative AI, Conversational AI, Transformer Models, Autonomous Agents, AI, Multi-agent Systems (MAS), LLM-based Agents, Generative Agents, Autonomous Web Agent (AWA).}}

    \item \textbf{Pillar B: The Domain/Discipline (AI Agent Architecture \& SE Task)} \\
   \textcolor{black}{Keywords included: \textit{Agent-based Systems, Modular Architectures for AI Agent, AI Agent Design, Autonomous Agent System Design, Adaptive AI Architectures, Adaptive AI Taxonomy, Software Architecture, System Design, Code Generation, Automated Program Repair (APR), Bug-Fixing, Fault Localization, Software Testing, Code Review, Requirements Engineering, Software Maintenance, Refactoring, Code Snippet Adaptation, Code Intent Extraction, Simulation Testing, Security, Prompt Engineering.}}
\end{itemize}

\textcolor{black}{These pillars report the candidate terms produced by the extraction step, rather than a single search string applied uniformly across all databases. As some extracted SE-task terms were broader than the architectural focus of our research questions, not all candidate terms were included in the final queries. We selected the terms most relevant to FM-based agents and their architectures when constructing the executable queries. The query format was then adapted to the supported fields and operators of each digital library. The exact queries executed for each database are available in our replication package.}

\paragraph{\textbf{Automated Search:}} With the derived search string, we conducted the automated search across seven major digital libraries to ensure broad coverage of the relevant literature. The search was configured to include papers published between July 2017 and July 2025, covering the period since the emergence of modern transformer-based architectures\cite{vaswani2017attention}. This process retrieved 822 articles from the ACM Digital Library, 2,990 from IEEE Xplore, 909 from ScienceDirect, 4,456 from Web of Science, 5,019 from SpringerLink, 4,670 from arXiv, and 778 from DBLP. Using the above search strategy, we initially obtained 19644 articles that potentially relate to our research.

\begin{table}[htbp]
  \centering
  \caption{\textcolor{black}{Inclusion and Exclusion Criteria for Study Selection}}
  \label{tab:criteria}
  \begin{tabular}{p{0.5cm} p{7.5cm}}
    \hline
    \multicolumn{2}{l}{\textbf{Inclusion criteria}} \\
    \hline
    (1) & The article claims that a Foundation Model or Large Language Model based agent is used. \\
    (2) & The article claims that the study involves a Software Engineering (SE) task or discusses software architecture. \\
    \hline
    \multicolumn{2}{l}{\textbf{Exclusion criteria}} \\
    \hline
    (1) & Short articles whose number of pages is less than eight. \\
    (2) & Duplicate articles or similar studies with different versions from the same authors. \\
    (3) & The article is not a primary research paper (e.g., books, theses, keynotes, editorials, panels). \\
    (4) & The article is published in a workshop or a doctoral symposium. \\
    (5) & The article is a grey publication (e.g., a non-peer-reviewed pre-print) published before 2023. \\
    (6) & The article is not written in English. \\
    (7) & The article mentions the use of FM-based agents without describing the employed techniques or architecture. \\
    (8) & Studies where the primary contribution is the empirical evaluation of an agent's performance on an SE task, rather than a novel contribution to the agent's architectural design, patterns, or principles. \\
    \hline
  \end{tabular}
\end{table}

\textcolor{black}{\paragraph{\textbf{Exclusion \& Inclusion Criteria:}} Next, we further evaluated the relevance of these articles based on the inclusion and exclusion criteria shown in Table~\ref{tab:criteria} \cite{naveed2024model, wang2022machine}. The article selection process, as illustrated in Figure~\ref{fig:methodology}, consists of several phases. In the first phase, we conducted an automated filtering to exclude articles with fewer than eight pages (Exclusion Criterion 1), reducing the number of articles to 13991.} \textcolor{black}{In the second phase, we examined the titles and abstracts of the articles to identify those that directly relate to our research topic. The purpose of this phase was to apply our core inclusion criteria, ensuring that the studies discuss both FM-based agents and their application or architecture in software engineering (Inclusion Criteria 1, 2). We also excluded non-English literature during this phase (Exclusion Criterion 6). After the second phase, the number of articles was reduced to 4841.}

The third phase involved identifying the venues and types of articles. We excluded studies belonging to books, theses, editorials, and other non-primary sources (Exclusion Criterion 3), as well as those published in workshops or doctoral symposiums (Exclusion Criterion 4). We also applied a filter for grey literature, such as non-refereed technical reports (Exclusion Criterion 5). We extracted publication information such as “journal,” “URL,” “DOI,” and “series” to determine the publication sources. For articles from arXiv from 2023 to 2025, we decided to keep them, since this field is emerging and many works are in the process of submission. To mitigate the risks of including non-peer-reviewed work from 2023-2025, we implemented a stringent quality assessment (QA) protocol (Table~\ref{tab:qaqs}). This step resulted in 1851 articles. In the fourth phase, we merged and deduplicated the remaining articles from the seven databases (Exclusion Criterion 2), resulting in 1588 unique articles. We then reviewed the full texts of these articles. This phase primarily involved excluding articles that mentioned FM-based agents only peripherally or without describing the employed techniques and architecture (Exclusion Criteria 7, 8). By further assessing the quality of the articles, we identified 640 articles directly relevant to our research. \textcolor{black}{GPT-4o was used only as an assistive tool during screening. At each screening round, the primary researcher reviewed the model-generated filtering results before proceeding, and all final inclusion and exclusion decisions were made by the researcher.} To ensure transparency, all prompts, and raw filtering records are provided in our replication package for verification. To ensure reliability, a second senior researcher audited the inclusion/exclusion decisions for the candidate set of 640 papers. This process achieved an agreement rate of 92\% (calculated specifically on the candidate set, excluding papers previously filtered by GPT-4o), with any disagreements resolved through discussion.

\textcolor{black}{\paragraph{\textbf{Quality Assessment:}} To ensure the reliability and validity of the selected primary studies and to mitigate potential biases introduced by low-quality research, we conducted a formal quality assessment (QA)~\cite{hou2024large}. This process is a standard practice in systematic literature reviews. We designed a checklist of 10 Quality Assessment Questions (QAQs), shown in Table~\ref{tab:qaqs}, inspired by established SLR guidelines and tailored to the specific goals of our research on agent architectures.} \textcolor{black}{The assessment was conducted in a two-stage process. In the first stage, QAQ1 and QAQ2 served as a final check on our inclusion criteria. Each question was answered with a simple \textbf{Yes (1 point)} or \textbf{No (0 points)}. A study had to receive a score of 1 or 2 to proceed to the second stage of quality scoring; if the score was 0, it was excluded. In the second stage, we evaluated the quality and rigor of the remaining studies using QAQ3 through QAQ10. Each of eight questions was scored on a three-point scale: \textbf{Fully Met (2 points)}, \textbf{Partially Met (1 point)}, or \textbf{Not Met (0 points)}. To account for the inclusion of recent pre-prints from arXiv, which is a common practice in this fast-evolving field, we applied a differentiated scoring threshold:}

\begin{itemize}
    \item \textbf{For peer-reviewed articles:} \textcolor{black}{All eight questions (QAQ3--QAQ10) were scored, leading to a maximum possible score of 16. We included studies that achieved a total score of \textbf{12 or higher} (approximately 75\%).}
    \item \textcolor{black}{\textbf{For non-peer-reviewed pre-prints (e.g., from arXiv):} QAQ9 was treated as not applicable because it assesses peer-review status and venue standing, which cannot yet be observed for a pre-print. The score was therefore calculated using the remaining seven questions, resulting in a maximum score of 14. Pre-prints were included if they achieved a score of at least 10 out of 14 (approximately 71.4\%).}
\end{itemize}

\textcolor{black}{For QAQ9, venue standing was assessed using recognised indicators relevant to the venue type, including CORE rankings for conferences, JCR or SJR rankings and impact indicators for journals, and the recognised standing of the venue within the software engineering and AI research communities. A score of 2 was assigned to an established peer-reviewed venue, a score of 1 to a peer-reviewed venue with limited or emerging evidence of established standing, and a score of 0 when the venue’s peer-review status or standing could not be verified.}

\textcolor{black}{QAQ9 assesses peer-review status and venue standing rather than the methodological quality of the study itself. For pre-prints, this dimension was treated as not applicable because the peer-review and publication process had not been completed at the time of data collection. Following Hou et al. \cite{hou2024large}, their quality scores were therefore calculated using the remaining seven questions. This treatment does not imply that a pre-print is equivalent to a paper published in a reputable venue; it only indicates that the venue-related dimension could not yet be assessed.}

This structured and transparent scoring rubric ensures that all included studies, regardless of their publication status, meet a consistent and high standard of quality and relevance to our research. \textcolor{black}{Specifically, 47 papers were excluded at Stage 1 (QAQ1-QAQ2) and 538 at Stage 2 (scoring). The scoring was performed by a primary researcher and audited by a senior researcher, with any discrepancies resolved through negotiated agreement. The filtering records for both stages and the detailed scoring for the final studies are provided in our replication package.} After this quality assessment, a final set of 55 primary studies were selected for data extraction.

\begin{table}[t]
\caption{\textcolor{black}{Checklist of Quality Assessment Questions (QAQs)}}
\label{tab:qaqs}
\small
\begin{tabular}{lp{7cm}} 
\toprule
\textbf{ID} & \textbf{Quality Assessment Question} \\
\midrule
\multicolumn{2}{l}{\textit{\textbf{--- Stage 1: Final Relevance \& Scope Check ---}}} \\
QAQ1 & Does the study definitively propose, implement, or analyze an agent based on Foundation Models (FMs) or Large Language Models (LLMs)? \\
QAQ2 & Is the study a primary research paper (i.e., not a survey, systematic review, tutorial, or editorial)? \\
\midrule
\multicolumn{2}{l}{\textit{\textbf{--- Stage 2: In-depth Quality \& Rigor Assessment ---}}} \\
QAQ3 & Is the motivation for the research and the problem statement clearly defined? \\
QAQ4 & Does the study provide a clear and understandable description of the agent's architecture \\
     & (e.g., its components, connectors, and their relationships)? \\
QAQ5 & Is the research methodology (e.g., case study, experiment, empirical study) clearly \\
     & described and appropriate for the research questions? \\
QAQ6 & Does the study include an evaluation or a detailed discussion of the agent's capabilities \\
     & or the trade-offs of its architectural choices? \\
QAQ7 & Is there a clear statement of findings? \\
QAQ8 & Is the research novel and does it contribute to the field? \\
QAQ9 & Was the paper published in a reputable, peer-reviewed venue (e.g., a well-known \\
     & conference or journal)? \\
QAQ10 & Are the limitations of the study or threats to validity clearly described? \\
\bottomrule
\end{tabular}
\end{table}

\paragraph{\textbf{Snowballing:}} To ensure the comprehensiveness of our study and to identify any relevant primary studies that may have been missed during the database search, we conducted a snowballing search process. Snowballing is a systematic technique that involves using the reference list of an article (backward snowballing) and the citations to an article (forward snowballing) to identify additional relevant literature~\cite{usman2017taxonomies}. The starting point for this process was the final set of 55 primary studies that passed our quality assessment. We performed both backward and forward snowballing on this entire set. The backward pass, which involved scanning the reference lists of our included studies. The forward pass involved identifying all articles that cited our included studies (using Google Scholar's "Cited by" feature). After merging and deduplicating these two sets, this new set of articles was then subjected to the exact same multi-phase screening and quality assessment process described in the above steps. This rigorous re-evaluation resulted in the inclusion of an additional 11 primary studies. \textcolor{black}{By combining the results from our automated database search and the snowballing search, the final set of primary studies selected for data extraction and synthesis consists of \textbf{66} articles (\textbf{52} peer-reviewed articles and \textbf{14} pre-prints)}.

\subsubsection{B8: Perform Terminology Control}
\textcolor{black}{During the data extraction phase (B7), it became evident that the primary studies used a diverse and often inconsistent set of terms to describe similar architectural concepts, a common challenge in a rapidly evolving field like FM-based agents. To ensure the conceptual integrity and prevent ambiguity in our taxonomy, a systematic Terminology Control step (B8) was essential~\cite{usman2017taxonomies}. The goal of this step was to systematically harmonize synonymous or overlapping terms into a single, consistent concept that would be used throughout our taxonomy.}

\textcolor{black}{Our process involved using the data extraction form to explicitly document the original terminology used in each paper alongside the architectural concept it described. Through an iterative process of thematic analysis and consensus among the research team, these source terms were mapped to a standardized set of categories. A prime example of this process can be seen in the dimension of \textit{Agent Coordination}. Our review of the literature revealed multiple terms describing architectures where a single agent manages or directs others. For instance, we encountered phrases such as \textit{`orchestrator model'}, \textit{`hierarchical control'}, \textit{`master-slave pattern'}, and \textit{`central controller'}. These were all systematically unified under the single, standardized category of \textbf{Centralized} coordination within our taxonomy. Similarly, for coordination patterns without a central authority, terms like \textit{`peer-to-peer (P2P) interaction'}, \textit{`decentralized consensus'}, and \textit{`emergent collaboration'} were identified and harmonized into the category of \textbf{Fully Distributed} coordination.}

\textcolor{black}{This rigorous process of terminology control was applied to all concepts extracted from the literature, ensuring that the final taxonomy is built upon a clear, consistent, and well-defined conceptual foundation.}

\subsubsection{B9: Identify and Describe Taxonomy Dimensions}
Following terminology control (B8), the next step was to identify the taxonomy's high-level dimensions from the synthesized data. This was not a pre-determined step but an emergent process driven by the evidence from the primary studies. We employed a thematic analysis approach, using an iterative process of open and axial coding on all extracted architectural features to group related concepts into higher-level themes. This bottom-up analysis led to the emergent definition of the \textbf{Functional Capabilities} dimension. For the \textbf{Non-functional Qualities} dimension, we adopted a hybrid approach. \textcolor{black}{We began with a top-down synthesis of candidate qualities from established standards, including \textbf{ISO/IEC 25010} \cite{iso25010} and the foundational work \textit{Software Architecture in Practice} \cite{bass2021architecture} and \textit{Engineering AI Systems} \cite{bass2025engineering}.} We utilized these definitions as an initial codebook, performing a systematic deductive mapping to validate and refine them against the architectural evidence extracted from our primary studies. This process resulted in the two primary, orthogonal dimensions of our taxonomy, which are formally defined as:

\begin{itemize}
    \item \textcolor{black}{Functional Capabilities: Consolidates all architectural options that define an agent's core operational mechanisms (e.g., how it perceives, reasons, and acts).} 

    \item \textcolor{black}{Non-functional Qualities: Groups the system-level quality attributes (e.g., reliability, performance) that constrain or define the overall quality of the architecture.}
\end{itemize}

\textcolor{black}{This data-driven derivation confirms the initial scope defined in B3 and provides an empirically grounded structure for the subsequent classification of categories in step B10.}

\subsection{Design \& Construction}
\label{subsec:design-construction}
\textcolor{black}{Following the establishment of the taxonomy's foundational dimensions, the subsequent steps focused on the detailed construction and rigorous validation of the classification scheme, adhering to the procedures B10 through B13 of the adopted methodology~\cite{usman2017taxonomies}.}

\begin{itemize}
    \item \textcolor{black}{\textbf{B10 \& B11: Identify Categories and Their Relationships.} For each high-level dimension, specific categories representing distinct architectural options were identified through a systematic synthesis of the extracted data. Our design aimed for orthogonality at multiple levels of the taxonomy. Not only are the two primary dimensions (Functional Capabilities and Non-functional Qualities) conceptually independent, but the sub-dimensions within each category (e.g., Memory Management, Planning, and Action) are also designed to be mutually exclusive and non-overlapping. \textcolor{black}{We acknowledged that strict mutual exclusivity is challenging in this domain; therefore, we resolved borderline cases by classifying components based on their primary architectural intent within the specific system context.} This structural integrity was formally verified through our expert-driven validation process, as detailed in Section~\ref{sec:validation}. The complete taxonomy is presented in Section~4.}

     \item \textcolor{black}{\textbf{B12: Define Guidelines for Using the Taxonomy.} To translate the descriptive taxonomy into prescriptive guidance for architects, we developed a decision model. This model is presented as a Decision Matrix in Section~\ref{sec:decision_model} and explicitly maps architectural choices to their impact on system qualities, thereby guiding practitioners through critical design trade-offs.}
\end{itemize}

\subsection{Validation}
\textcolor{black}{While the rigorous SLR process ensured the traceability and groundedness of the taxonomy's construction, we relied on a formal, external validation as our primary validation mechanism to assess the taxonomy's completeness, correctness, and practical relevance. As detailed in Section~\ref{sec:validation}, this external validation involved an expert-driven assessment using a structured questionnaire with participants from both industry and academia.}

\textcolor{black}{The replication package for this study, including the detailed search terms, the full list of screened studies with decisions, raw extracted data, expert validation protocol and results, and analysis scripts, is publicly available at\footnote{https://github.com/submissionpurposeonly/jss/tree/main}.}

\section{\textcolor{black}{Taxonomy Reporting and Validation}}
\label{sec:taxonomy}

\begin{figure*}
    \centering
    \includegraphics[width=0.6\linewidth]{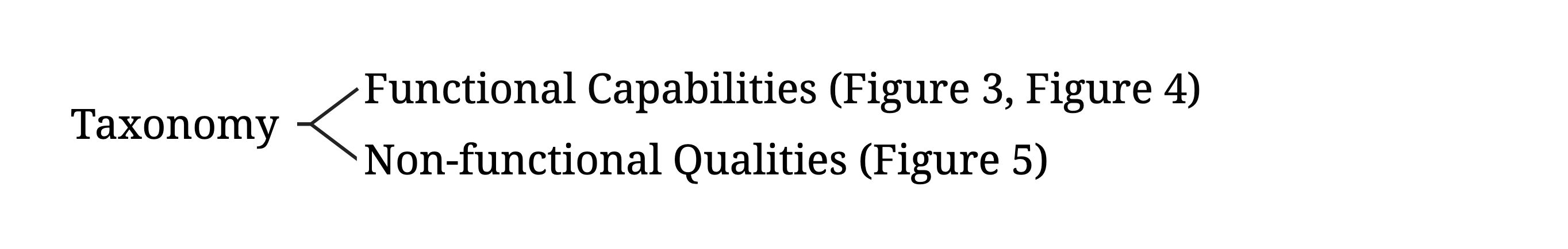}
    \caption{The high-level structure of the taxonomy. Two primary dimensions: Functional Capabilities and Non-functional Qualities.}
    \label{fig:taxonomy_overview}
\end{figure*}

\begin{figure*}
    \centering
    \includegraphics[width=0.65\linewidth]{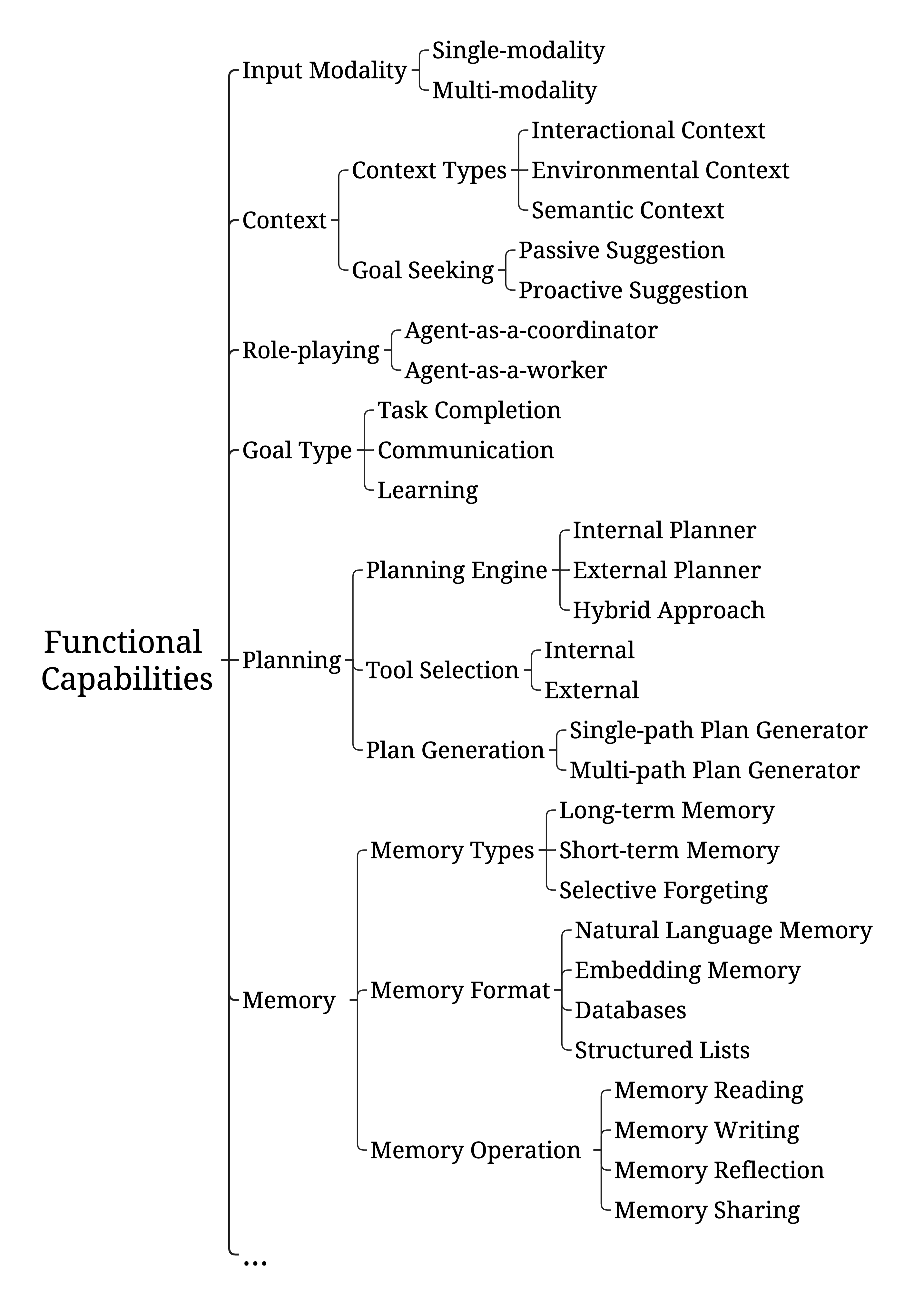}
    \caption{The taxonomy of Functional Capabilities (Part 1 of 2), detailing architectural options from Input Modality to Memory.}
    \label{fig:taxonomy01}
\end{figure*}

\begin{figure*}
    \centering
    \includegraphics[width=0.88\linewidth]{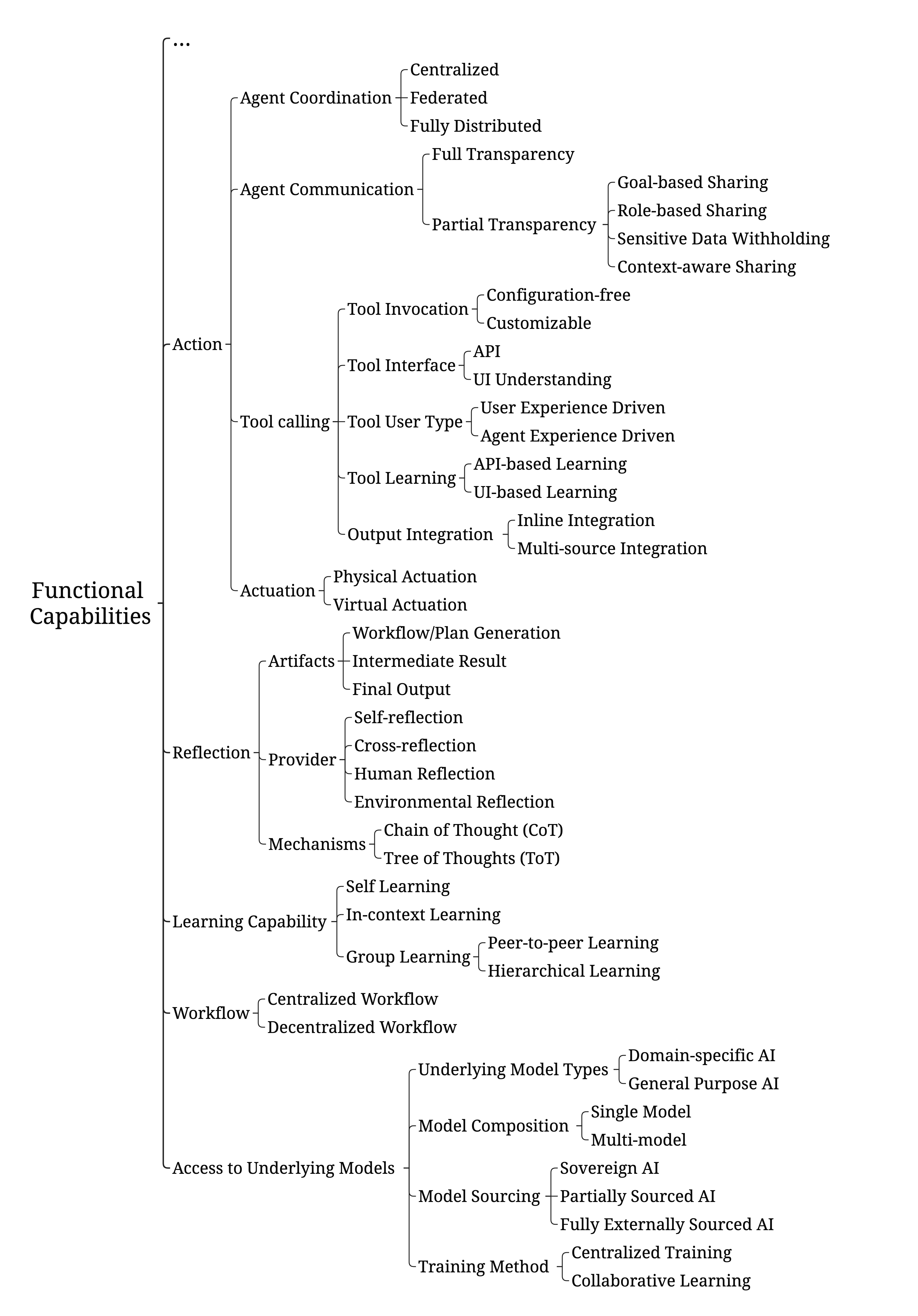}
    \caption{The taxonomy of Functional Capabilities (Part 2 of 2), detailing architectural options from Action to Access to Underlying Models.}
    \label{fig:taxonomy02}
\end{figure*}

\begin{figure}
    \centering
    \includegraphics[width=0.9\linewidth]{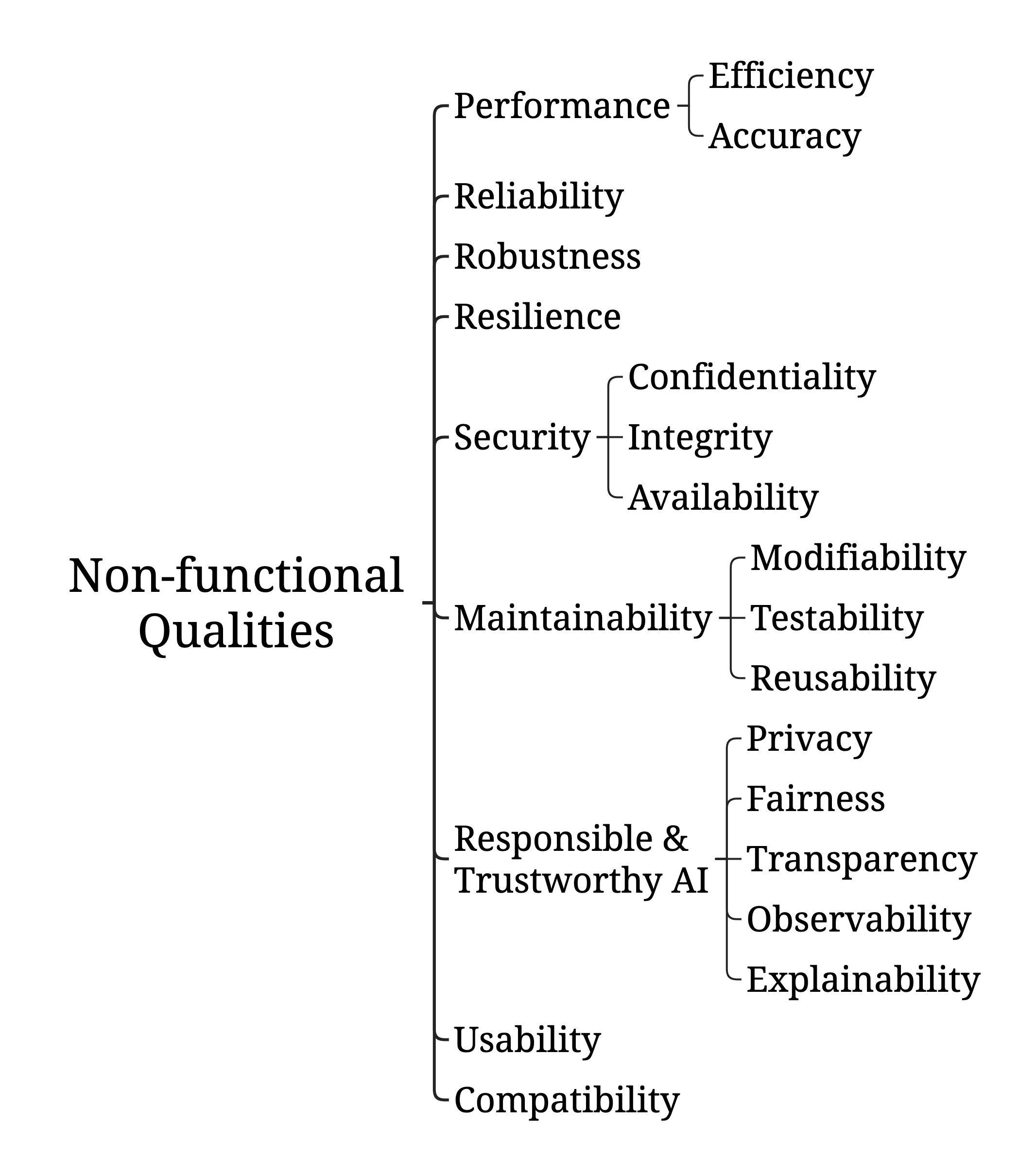}
    \caption{Taxonomy - Non-functional capabilities of foundation-model-based agent}
    \label{fig:nonfunctional}
\end{figure}

\textcolor{black}{In this section, we present a taxonomy that defines architecture design options for FM-based agents. The taxonomy is structured into two categories: functional capability (Section~\ref{subsec:input_modality} - Section~\ref{subsec:undermodel}) and non-functional qualities (Section~\ref{sec:non-functional}). Figure~\ref{fig:taxonomy_overview} illustrates the high-level structure of this taxonomy, while detailed breakdowns of the functional capabilities are presented across Figure~\ref{fig:taxonomy01} and Figure~\ref{fig:taxonomy02}, and the non-functional qualities are shown in Figure~\ref{fig:nonfunctional}. We discuss the characteristics of each design option to consider during the building process and their impact on the FM-based agents. Additionally, we provide a mapping table~\ref{tab:mapping_table} that maps each of our 66 primary studies to the corresponding taxonomy categories.}

\subsection{\textcolor{black}{Input Modality}}
\label{subsec:input_modality}
Modality defines whether an agent operates using a single type of data input or multiple. \textbf{Single-modality} agents process one type of input, such as text, vision, or audio \cite{touvron2023llama}. This approach is well-suited for straightforward tasks where computational efficiency and simpler data interpretation are priorities, as seen in text-based chatbots or vision-only surveillance systems \cite{xia2023towards}. In contrast, \textbf{multi-modality} agents are architected to integrate and process a combination of inputs---including text, audio, images, and video---to gain a more holistic understanding of their operational environment \cite{chang2025drc, you2025ferret}. For instance, by processing simultaneous speech and visual gestures, these agents can provide more accurate and context-aware responses \cite{xu2024fedmllm}. The capability to synthesize diverse data streams is a critical architectural driver for complex applications, such as autonomous navigation or advanced interactive assistants, which demand high levels of adaptability and contextual awareness.

\subsection{\textcolor{black}{Context}}
\label{subsec:context-management}
Context management focuses on gathering and organizing information about the agent's operating environment to understand user intentions and goals better. Our taxonomy classifies this into two sub-dimensions: the types of context gathered and the strategies for goal seeking. The classification of \textbf{context types} is based on the nature of the information itself. \texttt{Interactional context} refers to direct user actions and behaviors, such as mouse clicks, typing patterns, and gestures \cite{zhao2023seehow}. \texttt{Environmental context} describes the state of the digital environment in which the user is operating \cite{yang2024generalized}. Finally, \texttt{Semantic context}, extracted from sources like user annotations or prompt content, pertains to the underlying meaning and intent, providing crucial insights for goal interpretation \cite{modarressi2023ret}. For \textbf{goal seeking}, agents can employ two primary strategies. With \texttt{passive suggestion}, the agent analyzes goals that are explicitly articulated by the user, typically through a dialogue interface \cite{you2025ferret}. In contrast, \texttt{proactive suggestion} goes beyond explicit commands, enabling the agent to anticipate user goals by interpreting multimodal context from UI elements and user interactions \cite{gu2024middleware}. The architectural choice between these strategies presents a fundamental trade-off: passive suggestion offers implementation simplicity, whereas proactive suggestion enhances the agent's autonomy and intelligence at the cost of greater system complexity.

\subsection{\textcolor{black}{Role-Playing}}
\label{subsec:role-playing}
In multi-agent systems, assigning distinct roles is a crucial architectural decision that defines the functions and interactions within the system. The taxonomy categorizes these roles into two primary types: \textbf{agent-as-a-coordinator} and \textbf{agent-as-a-worker}. An \textbf{agent-as-a-coordinator} primarily formulates high-level strategies and orchestrates task execution by delegating responsibilities to other agents or external systems, ensuring the entire system operates cohesively and efficiently \cite{hong2023metagpt, lu2024towards, wu2023autogen, yao2022react}. In contrast, an \textbf{agent-as-a-worker} acts as the core executor. Upon receiving assignments, it is responsible for generating local, sub-task level strategies to complete them \cite{chang2025drc, gandhi2024budgetmlagent, islam2024mapcoder, mostajabdaveh2024optimization, takagi2025framework, xu2025mantra}. To achieve its objectives, a worker agent may operate with varying degrees of autonomy, interact with other agents through cooperation or competition, and call external tools. This architectural pattern enables a clear separation of concerns, allowing for strategic oversight by coordinators while leveraging the specialized, tactical execution of workers to solve complex problems.

\subsection{\textcolor{black}{Goal Type}}
\label{subsec:goal-type}
Agents establish goals to guide their planning and action phases, which our taxonomy classifies into three primary types: task completion, communication, and learning. \textbf{Task completion} goals direct an agent to achieve specific, complex objectives. Examples include agents programmed to craft items in virtual environments like Minecraft \cite{chen2023agentverse, wang2023voyager} or to execute predefined functions during software development \cite{qian2023communicative}. \textbf{Communication} goals involve agents interacting with other agents or humans to exchange information or collaborate on joint tasks. For instance, agents within platforms like ChatDev coordinate their efforts on development tasks \cite{qian2023communicative}, while others are designed to adapt their strategies based on real-time human feedback. Finally, \textbf{learning} goals require agents to navigate and adapt to unfamiliar settings, balancing the exploration of new areas against the exploitation of known resources. The Voyager agent exemplifies this, as it explores and refines its skills through continual feedback and adjustment processes \cite{wang2023voyager}.

\subsection{Planning}
\label{sec:planningprocess}
\textcolor{black}{In FM-based agents, planning is operationalized through a reasoning process, where the agent uses cognitive steps and logical frameworks to solve complex problems \cite{islam2024mapcoder, zhang2024codeagent}. This process bridges perception and action by enabling informed decision-making, which is implemented through architectural choices in how the agent selects tools and generates plans.}

\subsubsection{\textbf{Planning Engine}}
The planning engine is the central component orchestrating high-level strategic activities. It enables the agent to process reasoning, generate and monitor plans, and adapt to environmental shifts \cite{liu2024position}. Architectural options include:
\begin{itemize}
    \item \textbf{Internal Planner:} Embedded within the agent's core architecture, this planner uses the agent's intrinsic capabilities to autonomously generate and execute plans \cite{hao2023reasoning}.
    \item \textbf{External Planner:} Also known as classical planners, these are specialized, often domain-specific tools integrated to handle complex tasks requiring detailed and precise planning logic \cite{guan2023leveraging, valmeekam2409llms}.
    \item \textbf{Hybrid Approach:} This approach combines an internal FM-based planner for flexible problem translation with a classical external planner for efficient and reliable execution, leveraging the strengths of both \cite{valmeekam2409llms}.
\end{itemize}

\subsubsection{\textbf{Tool Selection}}
During planning, an agent must select the appropriate tools to execute its plan. The architectural choice lies between using tools that are internal to the agent's design or external to it.
\begin{itemize}
    \item \textbf{Internal Tools:} These are capabilities and algorithms built directly into the agent's system to provide tailored solutions. Examples include proprietary LLMs for reasoning, or specialized, self-contained algorithms and knowledge bases, such as those developed for the Voyager agent to manage its skills \cite{qu2024tool, wang2023voyager}.
    \item \textbf{External Tools:} These are outside services and data sources that the agent interacts with, typically via APIs. This allows the agent to extend its capabilities beyond its inherent functions, for instance, by calling a web search API, a code interpreter, or other external computational models to acquire new information or perform specialized tasks \cite{qu2024tool}.
\end{itemize}

\subsubsection{\textbf{Plan Generation}}
Plan generation translates the reasoning process into a sequence of actionable steps for the agent to follow \cite{ye2023proagent}. There are two primary architectural options for this process:
\begin{itemize}
    \item \textbf{Single-path Plan Generator:} This approach creates a linear, step-by-step plan. It is ideal for tasks in predictable environments where a straightforward procedure is effective, ensuring an efficient and direct route to the goal \cite{wang2023rcagent, zhou2309agents}.
    \item \textbf{Multi-path Plan Generator:} In contrast, this approach devises several potential routes to achieve a goal. It allows the agent to dynamically adjust its plan based on new information, making it more flexible and robust in complex or uncertain scenarios \cite{hao2023reasoning, wang2406mixture, wang2022self, yao2024tree}.
\end{itemize}
The choice between these generators represents a critical architectural trade-off between the efficiency and predictability of single-path plans and the adaptability and resilience of multi-path plans.

\subsection{\textcolor{black}{Memory}}
\textcolor{black}{Memory is a critical architectural dimension that enables FM-based agents to store, retrieve, and utilize information from past interactions to inform current and future actions. An agent's ability to maintain context, learn from experience, and exhibit consistent behavior over time is fundamentally dependent on the design of its memory system. This section details the architectural options related to memory, including its types, the formats used for storing information, and the core operations for manipulating it.}

\begin{itemize} 
    \item \textbf{Memory Types:}  
    \textcolor{black}{\textbf{Long-term memory} retains information over long periods, which is essential for tasks that require historical data, knowledge, past observations, and past experiences \cite{maharana2024evaluating, zhou2309agents}. \textcolor{black}{\textbf{Short-term memory}, on the other hand, handles short-term information relevant to immediate tasks \cite{park2023generative}.} This type of memory is useful for temporary activities that do not require long-term storage, for example, configuration, working context, recent events, and the information within the context window of the FM.} \textcolor{black}{\textbf{Selective forgetting} enables agents to discard irrelevant or outdated information, ensuring that they maintain optimal performance without being bogged down by unnecessary data \cite{park2023generative}.} This capability is crucial in dynamic environments where the relevance of data changes rapidly. 
    \item \textbf{Memory Format:} In the FM-based agent systems, several distinctive structures offer unique advantages and serve as pivotal design options depending on specific application needs. \textbf{Natural language memory}, as used in systems like Reflexion \cite{shinn2024reflexion} and Voyager \cite{wang2023voyager}, provides a flexible format that stores information in an easily understandable form, facilitating intuitive interaction and preserving rich semantic details to guide agent actions. \textbf{Embedding memory}, exemplified by MemoryBank and ChatDev, encapsulates memory information into compact embedding vectors, significantly enhancing the efficiency of memory retrieval processes \cite{zhong2024memorybank}. \textbf{Databases} allow for robust memory manipulation; systems like ChatDB \cite{qian2023communicative} and DB-GPT use databases to enable precise memory operations through SQL queries, providing structured and efficient data management. \textcolor{black}{\textbf{Structured lists} are employed in models such as GITM and RET-LLM, where memory is systematically organized in lists or hierarchical structures that clearly define the relationships between elements and facilitate rapid data access \cite{park2023generative, modarressi2023ret}.} Each format presents distinct advantages: natural language for clarity and semantic richness, embeddings for retrieval speed, databases for structured manipulation, and lists for organized and hierarchical memory storage. These memory formats can be integrated, as demonstrated by GITM, which uses a hybrid approach combining key-value pairs with embeddings and natural language to maximize retrieval efficiency and content comprehensiveness, providing multiple design options to optimize agent performance based on the specific needs of the application.

    \item \textbf{Memory Operation:} Agents can acquire, accumulate, and utilize knowledge through interactions with their environment via various operations. These operations include memory reading, writing, reflection, and sharing. \textcolor{black}{\textbf{Memory reading} involves extracting valuable information based on recency, relevance, and importance to enhance the agent’s actions \cite{gandhi2024budgetmlagent, park2023generative, xu2025mantra, islam2024mapcoder}.} \textbf{Memory writing} focuses on storing perceived environmental information while managing duplication and preventing overflow \cite{achiam2023gpt}. \textbf{Memory reflection} enables agents to summarize and infer high-level insights from past experiences, facilitating more abstract and complex decision-making \cite{park2023generative}. \textcolor{black}{\textbf{Memory sharing} allows agents to access and contribute to a common memory pool. This enhances their collaborative abilities by integrating diverse experiences and knowledge \cite{maharana2024evaluating, zhang2024survey, mei2024aios, gao2024memory}.} In FM-based agent systems, memory sharing involves storing and retrieving shared memories, which supports continuous learning and adaptation, helping agents use the most relevant information for tasks. 
\end{itemize}

\subsection{Action}
\label{subsec:action}
\textcolor{black}{The Action dimension encompasses the architectural choices that govern how an agent executes tasks and interacts with its environment, other agents, or external tools. It represents the phase where the agent translates its internal plans and decisions into tangible operations that effect change. This section describes the key architectural options within this dimension, including the mechanisms for agent coordination, the strategies for communication, the methods for tool calling, and the types of actuation.}

\subsubsection{\textbf{Agent Coordination}}
\label{sec:agentcoordination}
At runtime, agent coordination mechanisms ensure that multiple agents can work together effectively, avoiding conflicts and redundancies. The taxonomy identifies three primary coordination patterns:
\begin{itemize}
    \item \textbf{Centralized Coordination:} This pattern follows a strict command-and-control model where a single "manager" agent dictates specific tasks and steps to other agents, which have limited autonomy. The system's core intelligence is concentrated at the center \cite{guo2024large, mei2024aios}.
    \item \textcolor{black}{\textbf{Federated Coordination:} In this model, a central "orchestrator" is responsible for high-level tasks like assigning goals or aggregating results, but it grants high autonomy to individual agents to manage their own internal execution logic and complex sub-tasks \cite{gandhi2024budgetmlagent, islam2024mapcoder, wu2023autogen, xu2025mantra, zhang2025empowering}.}
    \item \textcolor{black}{\textbf{Fully Distributed Coordination:} This approach involves no central authority; all agents are peers that coordinate directly through peer-to-peer communication to reach a consensus or synchronize actions \cite{chen2023agentverse, guo2024large, park2023generative}.}
\end{itemize}
The selection among these patterns represents a fundamental architectural trade-off between the simplicity and efficiency of centralized control versus the resilience and autonomy inherent in distributed systems.

\subsubsection{\textbf{\textcolor{black}{Agent Communication}}}
Agent communication strategies dictate how agents share information to collaborate, representing a crucial aspect of multi-agent system design. The primary architectural choice concerns the level of transparency:
\begin{itemize}
    \item \textbf{Full Transparency:} All information is openly shared among agents to maximize cooperation and situational awareness \cite{gao2024memory, wang2406mixture, zhou2309agents}. While ideal for deeply collaborative tasks, this approach can introduce risks of information overload, privacy breaches, and security vulnerabilities \cite{zhu2024survey}.
    \item \textbf{Partial Transparency:} Information sharing is restricted to enhance security and efficiency. This is achieved through several patterns, including: \textbf{Goal-based sharing}, where information is exchanged only if relevant to specific goals \cite{guo2024large}; \textbf{Role-based sharing}, where access is determined by an agent's responsibilities \cite{guo2024large}; \textbf{Sensitive data withholding} to protect privacy \cite{agashe2023evaluating, zhu2024survey}; and \textbf{Context-aware sharing}, where the decision to share is adapted dynamically based on the system's state \cite{agashe2023evaluating, du2024survey, pesce2023learning}.
\end{itemize}
Architecting the communication strategy therefore requires balancing the need for information availability against the critical system qualities of security, privacy, and efficiency.

\subsubsection{\textbf{\textcolor{black}{Tool Calling}}}
Tool calling is a fundamental capability that enables agents to access and manipulate external tools. This dimension is composed of several key architectural choices.

\textbf{Tool Invocation} defines how an agent interacts with a tool. With \texttt{Configuration-free} invocation, agents use tools via predefined interfaces without adjustment, ideal for standardized, routine operations \cite{schick2023toolformer}. In contrast, \texttt{Customizable invocation} allows agents to dynamically adjust tool parameters to optimize performance for specific tasks \cite{deng2023mind2web, qiao2023making, schick2023toolformer}. For example, a code generation agent can customize a \texttt{create\_function} tool by specifying parameters like \texttt{language='Python'} and \texttt{args=['user\_id']}, enabling both high precision for current requirements and adaptability for future changes.

\textbf{Tool Interface} is the layer through which an agent operates a tool. An \texttt{API} provides a structured, machine-to-machine interface, enabling reliable and efficient interaction with services that expose a programmatic endpoint \cite{durante2024interactive, feng2024prompting, jimenez2023swe, qu2024tool, schick2023toolformer, xi2023rise}. \textcolor{black}{Alternatively, \texttt{UI Understanding} allows an agent to operate a tool by perceiving and interacting with its user interface, mimicking human behavior \cite{deng2023mind2web}. This is crucial when no API is available \cite{baechler2024screenai} and can be applied to multimodal interfaces, including visual, audio, or gesture-based systems \cite{baechler2024screenai, zhang2024large}.}

\textbf{Tool User Type} determines the source of feedback for tool refinement. \texttt{User experience-driven} tools evolve based on direct human interaction and feedback, prioritizing usability and user satisfaction \cite{brand2023towards, GoogleAI2024}. \texttt{Agent experience-driven} tools, on the other hand, harness the agent's own historical performance data to autonomously refine and optimize operational strategies \cite{liu2024position}. This choice represents an architectural trade-off between optimizing for human-centric usability and for agent-centric autonomous performance.

\textbf{Tool Learning} describes how agents acquire new tool-use capabilities. In \texttt{API-based learning}, agents learn to master stable, structured interfaces, leading to highly reliable and efficient tool use \cite{feng2024prompting, qu2024tool, schick2023toolformer}. In \texttt{UI-based learning}, agents learn to interpret and adapt to potentially inconsistent or evolving user interfaces, which grants greater flexibility at the cost of potential brittleness \cite{deng2023mind2web, you2025ferret}.

\textbf{Output Integration} is the process of incorporating tool outputs into an agent's workflow. \texttt{Inline integration} involves directly using a tool's output for immediate task execution \cite{qu2024tool, schick2023toolformer}. \texttt{Multi-source integration} is a more complex approach where outputs from multiple tools are synthesized, often with the agent's internal data, to produce more nuanced and comprehensive results \cite{durante2024interactive}.

\textbf{Actuation} refers to an agent's capability to execute actions that affect a physical or digital environment \cite{chanharms}. The taxonomy distinguishes between two types:
\begin{itemize}
    \item \textbf{Physical actuation} involves actions in the real world, such as the robotic movements and manipulations required for factory assembly tasks or the coordinated steering, braking, and acceleration in an autonomous vehicle \cite{gupta2024essential, yang2023foundation, yang2024generalized}.
    \item \textbf{Virtual actuation} encompasses actions within software environments, including automated data entry, network configuration changes, or interacting with software applications \cite{deng2023mind2web, huang2025comprehensive, jimenez2023swe, liu2023agentbench}.
\end{itemize}

\subsection{\textcolor{black}{Reflection}}
\label{subsec:reflection}

\textcolor{black}{Reflection is the architectural dimension that equips an agent with meta-cognitive capabilities to introspectively analyze its own performance and behavior in order to learn and improve. This process of self-assessment is crucial for enabling adaptive and robust agents that can correct their mistakes and refine their strategies over time. This section details the architectural options for reflection, including the artifacts that are reflected upon, the providers of feedback, and the specific mechanisms used for reasoning.}

\subsubsection{\textbf{Reflected Artifacts}}
Reflection can target different artifacts generated during an agent's operation to ensure traceability and enable feedback at various stages. The primary architectural options include:
\begin{itemize}
    \item \textbf{Workflow/Plan Generation:} This includes the agent's initial goals, prompts, and planning steps. Reflection at this stage assesses the validity and comprehensiveness of the agent's initial approach before execution \cite{longo2024explainable}.
    \item \textbf{Intermediate Result:} This artifact comprises the logs of reasoning processes, planning steps, and tool use during operation, allowing for reflection on the correctness and rationality of the agent's internal decision-making process \cite{mostajabdaveh2024optimization}.
    \item \textbf{Final Output:} This refers to the final result produced by the agent, which is evaluated against the initial goals to verify that the outcome meets the desired criteria \cite{mostajabdaveh2024optimization}.
\end{itemize}

\subsubsection{\textbf{Provider}}
Reflection is driven by feedback, which can originate from several providers \cite{park2023generative}. The architectural choice of provider determines the source of guidance for plan refinement:
\begin{itemize}
    \item \textbf{Self-reflection:} The agent generates its own feedback to evaluate and refine its plans, forming an internal loop of self-improvement \cite{chang2025drc, islam2024mapcoder, mostajabdaveh2024optimization, shinn2024reflexion, sumers2023cognitive, wang2023voyager, yao2022react}.
    \item \textbf{Cross-reflection:} One or more external agents or FMs are used to provide feedback, leveraging diverse perspectives to critique and enhance the plan \cite{shinn2024reflexion, wang2406mixture}.
    \item \textbf{Human Reflection:} The agent collects feedback directly from human users, which is crucial for aligning its behavior with human preferences and expectations \cite{guan2023leveraging, yang2023foundation, zhang2025empowering}.
    \item \textbf{Environmental Reflection:} The agent obtains feedback from the external world, such as task completion signals or observations after an action \cite{biagiola2024testing, jimenez2023swe, xu2025mantra, zhang2024codeagent}. For example, the ReAct framework uses thought-act-observation triplets for planning \cite{yao2022react}, while Voyager incorporates feedback from program execution outcomes and errors \cite{wang2023voyager}.
\end{itemize}

\subsubsection{\textbf{Mechanisms}}
Mechanisms are the specific techniques an agent employs to structure its reflection process. The primary architectural options are:
\begin{itemize}
    \item \textbf{Chain of Thought (CoT):} This mechanism involves the agent articulating its reasoning process step-by-step, which helps in tracing decisions and identifying errors \cite{feng2024prompting, wang2022self}. Advanced variants like CoCoNut extend this reasoning into a continuous latent space for greater efficiency \cite{hao2024training}.
    \item \textbf{Tree of Thoughts (ToT):} This more advanced mechanism allows an agent to explore multiple reasoning pathways simultaneously. By considering various possible outcomes before making a final decision, ToT enhances the robustness and depth of the agent's problem-solving capabilities \cite{hao2023reasoning, yao2024tree}.
\end{itemize}

\subsection{\textcolor{black}{Learning Capability}}
\label{sec:learning}   
Learning capabilities are fundamental to an agent's ability to adapt and improve its performance over time. The primary architectural options for enabling learning are:
\begin{itemize}
    \item \textbf{Self-learning:} This capability allows an agent to autonomously improve by updating its knowledge base and refining its decision-making algorithms. It often relies on long-term memory to learn from extensive historical data, enabling performance enhancement without direct external input \cite{maharana2024evaluating, peng2023self, shinn2024reflexion, wu2024copilot}.
    \item \textbf{In-context learning:} This enables an agent to adapt to new tasks on-the-fly by interpreting the provided context, typically through prompt engineering, without altering its internal model parameters \cite{gao2024memory, guan2023leveraging, hao2023reasoning, park2023generative, zhou2309agents}. This streamlined approach is highly effective for pre-trained models, allowing for rapid adaptation based on tailored prompts \cite{kamoi2024can}.
    \item \textbf{Group learning:} This involves multiple agents sharing knowledge and insights to enhance collective performance \cite{kamoi2024can, wang2406mixture}. It can be implemented through various architectural patterns, such as \textbf{peer-to-peer learning}, where agents share knowledge directly with each other \cite{gao2024memory, pesce2023learning}, or \textbf{hierarchical learning}, where information is processed and shared across different levels of an agent organization to solve complex problems more efficiently \cite{wang2406mixture}.
\end{itemize}

\subsection{\textbf{\textcolor{black}{Workflow}}}
The Workflow dimension governs the architectural pattern for the procedural execution of tasks. It is distinct from the strategic nature of \textit{Planning} (Section~\ref{sec:planningprocess}) and the interactive focus of \textit{Agent Coordination} (Section~\ref{sec:agentcoordination}), as its concern is task and control flow. The primary architectural designs are:
\begin{itemize}
    \item \textbf{Centralized Workflow:} Task management is concentrated at a single control point. This design enhances operational consistency but risks creating performance bottlenecks, a pattern analogous to centralized coordination \cite{mei2024aios}.
    \item \textbf{Decentralized Workflow:} Task control is distributed across multiple agents. This approach increases system flexibility and resilience, reflecting principles found in distributed coordination strategies \cite{chen2023agentverse}.
\end{itemize}

\subsection{Access to Underlying Models}
\label{subsec:undermodel}
\textcolor{black}{This dimension addresses the architectural choices related to the core engine of the agent: the foundation model(s) that provide its fundamental reasoning and generative capabilities. Decisions made in this dimension have a profound impact on the agent's core competence, cost, and operational constraints. This section details the key architectural options, including the types of underlying models, their composition, their sourcing, and the methods used for their training.}

\subsubsection{Underlying Model Types}
The choice of the underlying model type determines the agent's fundamental scope of competence. The primary options are:
\begin{itemize}
    \item \textbf{Domain-specific AI-based agents:} These agents are engineered to excel at a narrow set of specialized tasks within a particular domain, prioritizing high performance and accuracy over general applicability \cite{wu2023bloomberggpt}.
    \item \textbf{General Purpose AI (GPAI)-based agents:} In contrast, these agents are designed for versatility, capable of performing various tasks across different domains \cite{touvron2023llama}. They leverage broad training data and continuous learning mechanisms to adapt to new and unforeseen challenges, prioritizing flexibility over specialized expertise \cite{wu2024copilot}.
\end{itemize}

\subsubsection{Model Composition}
Model composition defines how many and what kinds of models constitute the agent's core engine. The architectural choice is between a single model or a multi-model configuration.
\begin{itemize}
    \item \textbf{Single-model:} The agent relies on a single, monolithic model to handle all of its tasks. This approach minimizes architectural complexity but may lack the scalability and specialized performance required for diverse tasks \cite{guo2024large}.
    \item \textbf{Multi-model:} The agent employs multiple models to enhance performance, robustness, or task-specific capabilities \cite{gandhi2024budgetmlagent, wang2406mixture, yang2024generalized}. This can be implemented through several advanced patterns, such as a \textbf{mixture of experts} for dynamic model selection \cite{yi2024fedmoe}, an \textbf{ensemble method} for improved accuracy \cite{lu2023routing}, \textbf{model merging} to create composite capabilities \cite{wu2024evolutionary}, or a \textbf{hybrid model} that combines different model types (e.g., rule-based and learning-based) to leverage their respective strengths \cite{kang2023grounding}.
\end{itemize}

\subsubsection{\textcolor{black}{Model Sourcing}}
Model sourcing addresses the strategic decision of how an agent's core AI technology is acquired, offering different trade-offs between control, cost, and customization. The options are:
\begin{itemize}
    \item \textbf{Sovereign AI:} The model is developed and maintained entirely in-house. This approach provides complete control over the model's capabilities and data security but requires significant investment \cite{touvron2023llama}.
    \item \textbf{Partially Sourced AI:} This hybrid approach combines internal development with external resources. For instance, BloombergGPT was developed by fine-tuning a base model on a unique mixture of proprietary financial data and public datasets to create a highly specialized, high-performance model \cite{wu2023bloomberggpt}.
    \item \textbf{Fully Externally Sourced AI:} The agent relies entirely on third-party AI models and services (e.g., via APIs), eliminating in-house development but ceding control over the model's architecture and evolution \cite{feng2024prompting}.
\end{itemize}

\subsubsection{\textcolor{black}{Training Method}} 
The training method is a critical architectural decision based on the system's data architecture—that is, whether training data is pooled centrally or remains distributed. The primary paradigms are:
\begin{itemize}
    \item \textbf{Centralized Training:} The traditional approach where a complete dataset is aggregated in a single location to train or fine-tune a model \cite{huang2025comprehensive, jimenez2023swe, touvron2023llama}. This ensures data consistency but can be constrained by resource requirements and privacy regulations.
    \item \textbf{Collaborative Learning:} This paradigm is used when data cannot be centralized, often due to privacy concerns \cite{gao2024memory}. It enables multiple parties to jointly train a model without sharing raw data, using patterns like \textbf{federated learning}, which shares model updates \cite{mabrouk2023ensemble}, which shares intermediate outputs from model layers.
\end{itemize}

While not parallel architectural choices in our taxonomy, two critical, cross-cutting concepts influence these paradigms and are essential for a complete architectural picture:
\begin{itemize}
    \item \textbf{Distributed learning} is a \textbf{computational strategy} focused on performance; it uses multiple machines to accelerate any large-scale training and can be applied to both centralized and collaborative approaches.
    \item \textbf{Online learning} refers to the \textbf{update frequency}, allowing an agent's model to be continuously refined in real-time as new data arrives, a choice orthogonal to whether the data architecture is centralized or collaborative.
\end{itemize}

\subsection{\textcolor{black}{Non-functional Qualities}}
\label{sec:non-functional}

\textcolor{black}{While functional capabilities define \textit{what} an agent does, non-functional qualities determine \textit{how well} it performs its functions. These qualities are critical system-level attributes that constrain the architectural design and dictate the overall quality and dependability of the agent. To provide a systematic and comprehensive classification, our taxonomy synthesizes principles from established software and AI systems architecture literature. \textcolor{black}{It is primarily grounded in the \textbf{ISO/IEC 25010 standard} \cite{iso25010} for software product quality, enriched with foundational quality attributes described by Bass et al. in \textit{Software Architecture in Practice} \cite{bass2021architecture} and \textit{Engineering AI Systems} \cite{bass2025engineering}.} \textcolor{black}{To ensure transparency and validity, we employed a deductive mapping procedure. We used the definitions from these standards as a predefined coding scheme. During data extraction, architectural concerns identified in the 66 primary studies were systematically mapped to these categories.} The key non-functional qualities relevant to FM-based agents are as follows:}

\paragraph{\textbf{Performance}} In the context of AI systems, performance is a multifaceted quality encompassing \textbf{Accuracy} and \textbf{Efficiency}. Accuracy measures the correctness of an agent's outputs, while Efficiency concerns the resources consumed (e.g., latency, computational cost, scalability) relative to the results achieved \cite{lu2023routing, wu2024evolutionary}.

\paragraph{\textbf{Reliability}} This refers to the degree to which an agent performs its specified functions correctly under normal conditions. High reliability implies stable operation (\textit{Maturity}), consistent service readiness (\textit{Availability}), and accurate results for in-distribution inputs \cite{chen2023agentverse, park2023generative}.

\paragraph{\textbf{Robustness}} Distinct from reliability, robustness is the agent's ability to maintain functionality in the presence of invalid, unexpected, or stressful inputs, such as noisy data or adversarial attacks—a critical quality for agents in unpredictable environments \cite{hao2023reasoning, yao2024tree}.

\paragraph{\textbf{Resilience}} Resilience is the ability of an agent to prepare for, absorb, recover from, and adapt to operational failures. It concerns the entire lifecycle of bouncing back from a significant failure, potentially to a degraded but still operational state \cite{guo2024large}.

\paragraph{\textbf{Security}} This quality addresses the protection of the agent, its data, and services. It encompasses: \textbf{Confidentiality} (protecting data from unauthorized access), \textbf{Integrity} (safeguarding assets from tampering), and \textbf{Availability} (ensuring services are accessible on demand) \cite{agashe2023evaluating, zhu2024survey}.

\paragraph{\textbf{Maintainability}} This concerns the ease with which an agent's architecture can be modified to correct faults, improve performance, or adapt. It is characterized by \textbf{Modifiability} (ease of change), \textbf{Testability} (ease of verification), and \textbf{Reusability} (leveraging components in new contexts) \cite{wang2023rcagent, zhou2309agents}.

\paragraph{\textbf{Responsible \& Trustworthy AI (RAI)}} This crucial cross-cutting concern integrates ethics and accountability into the agent's design. It includes characteristics such as \textbf{Privacy}, \textbf{Fairness}, \textbf{Transparency}, \textbf{Observability}, and \textbf{Explainability} (the ability to provide human-understandable justifications) \cite{longo2024explainable, lu2024towards}.

\paragraph{\textbf{Usability}} This quality addresses the effectiveness, efficiency, and satisfaction with which users can interact with an agent. It includes the intuitiveness of interaction, clarity of communication, and alignment with user intent \cite{brand2023towards, shinn2024reflexion, wang2023voyager}.

\paragraph{\textbf{Compatibility}} This quality refers to an agent's ability to exchange information and perform its functions while sharing an environment with other systems. It is vital for multi-agent systems and the integration of external tools via APIs \cite{durante2024interactive, qu2024tool}.

It is important to note that the \textbf{Level of Autonomy}, while a critical characteristic of an agent, is not classified here as a standalone non-functional quality. Instead, we consider it a crucial \textbf{emergent property} that arises from the specific combination of choices made within the \textbf{Functional Capabilities} dimension. An agent's ability to operate without human intervention is a direct consequence of its architectural design. Therefore, autonomy is evaluated as a result of the functional design, and its trade-offs against fundamental qualities like \textbf{Reliability} and \textbf{Maintainability} are best analyzed through a decision model, as we will demonstrate in Section~\ref{sec:decision_model_autonomy}.

\section{Decision Model for Architectural Trade-offs}
\label{sec:decision_model}
\textcolor{black}{The taxonomy presented in Section 4 provides a structured classification of architectural options that answers RQ1 and RQ2. However, it does not inherently guide an architect on \textit{which} option to choose in a given context to meet specific quality requirements. To bridge this gap between taxonomy and practical design, we introduce a decision model. This model aims to equip architects with a systematic tool to reason about the trade-offs inherent in designing FM-based agents.}

\subsection{Derivation of the Decision Model}
\textcolor{black}{The decision matrix is the result of a \textbf{systematic synthesis of the evidence collected from the primary studies} included in our SLR. The derivation process involved the following steps:}
\begin{itemize}
    \item \textcolor{black}{\textbf{Evidence Extraction}: During the data extraction phase of our SLR, we not only catalogued the architectural options used in each study but also explicitly extracted any claims, discussions, or empirical results related to the \textit{impact} of those choices on system-level qualities (e.g., statements about performance, reliability, maintainability).}
    \item \textcolor{black}{\textbf{Thematic Synthesis}: We then synthesized this extracted evidence. For each functional architectural choice in our taxonomy (e.g., \texttt{Multi-path Plan Generator}), we aggregated all associated non-functional impacts reported across the literature.}
    \item \textcolor{black}{\textbf{Impact Assessment}: Based on the consensus or prevalence of evidence in the literature, we assigned a qualitative impact assessment (\texttt{++}(strong positive), \texttt{+} (positive), \texttt{o}(neutral), \texttt{-}(negative), \texttt{--}(strong negative)) for each intersection in the matrix. For example, if multiple studies empirically demonstrated or strongly argued that a specific choice significantly enhances a quality, it was marked with \texttt{++}.}
\end{itemize}

\textcolor{black}{The impact ratings in the Decision Matrix were synthesized through a systematic analysis of the 66 primary studies. For each study, the primary researcher extracted observed design trade-offs and mapped them to non-functional qualities, recording these as qualitative observations in the data extraction repository (e.g., the "Non-functional Qualities" column in our data extraction form). To resolve instances of conflicting evidence, the primary researcher performed a cross-case comparison to identify the most prevalent architectural trends across the dataset. These synthesized ratings were then independently reviewed and cross-checked by a second researcher to ensure consistency and mitigate individual bias. Finally, the matrix was validated against the practical engineering scenarios included in our expert questionnaire (e.g., the "AI Analyst Agent" case), ensuring the model reflects a grounded consensus between synthesized literature evidence and practitioner experience.}

\subsection{The Architectural Trade-off Decision Matrix}
\textcolor{black}{ To illustrate how this matrix (Table \ref{tab:decision_matrix_final}) is used to navigate complex design trade-offs, we now analyze two foundational architectural scenarios drawn from the literature.}

\paragraph{\textbf{Architectural Scenario 1: The Planning Trade-off -- Flexibility vs. Predictability}}
This scenario concerns the choice of a plan generation strategy.
\begin{itemize}
\item \textbf{The Choice}: An architect must decide between a \texttt{Single-path Plan Generator} and a \texttt{Multi-path Plan Generator}.
\item \textbf{Positive Consequences}: Opting for a \texttt{Multi-path Plan Generator} endows the agent with high adaptability. As shown in our matrix, this greatly enhances \textbf{Usability} (\texttt{++}) in dynamic scenarios and improves \textbf{Robustness} (\texttt{++}) by allowing the agent to explore multiple solution pathways to overcome unexpected obstacles.
\item \textbf{Negative Consequences}: However, this flexibility introduces significant non-determinism. This makes the agent's behavior difficult to test and reproduce, which, as the matrix indicates, severely degrades \textbf{Maintainability} (\texttt{--}) and its sub-attribute, \textit{testability}.
\item \textbf{The Resulting Trade-off}: The architect is therefore forced into a classic trade-off: they must sacrifice the engineering predictability and ease of maintenance offered by a single-path approach in order to gain the high adaptability and robustness required for complex environments.
\end{itemize}

\paragraph{\textbf{Architectural Scenario 2: The Memory Trade-off -- Human-centric vs. Machine-centric Design}}
This scenario addresses the choice of memory format, a critical decision in designing explainable and efficient agents.
\begin{itemize}
\item \textbf{The Choice}: The decision lies between a human-readable format like \texttt{Natural Language Memory} and a machine-optimized format like \texttt{Embedding Memory}.
\item \textbf{Positive Consequences}: Choosing \texttt{Natural Language Memory} provides superior transparency. Its inherent human-readability, as reflected in the matrix, is crucial for \textbf{Responsible \& Trustworthy AI} (\texttt{++}) (specifically, explainability) and \textbf{Usability} (\texttt{++}), especially in domains requiring human oversight.
\item \textbf{Negative Consequences}: This transparency comes at a steep price. Processing unstructured text incurs high computational overhead, severely impacting \textbf{Performance} (\texttt{--}). Conversely, \texttt{Embedding Memory} offers highly efficient retrieval, boosting \textbf{Performance} (\texttt{++}), but its black-box nature is detrimental to both \textbf{RAI} (\texttt{--}) and \textbf{Usability} (\texttt{--}).
\item \textbf{The Resulting Trade-off}: The architect must make a conscious decision between prioritizing human-centric qualities like explainability and usability, or machine-centric qualities like computational performance and efficiency.
\end{itemize}

\subsection{A Composite Example: Architecting for Composite Goals}
\label{sec:decision_model_autonomy}
\textcolor{black}{Beyond analyzing the trade-offs of individual architectural choices, the decision model's primary utility lies in reasoning about \textbf{composite, real-world design goals}. These goals, such as achieving a high level of autonomy, are often not fundamental categories within the taxonomy itself, but are rather complex characteristics that emerge from the interplay of multiple architectural decisions. A prime example is architecting for a \textit{high level of autonomy}. As we defined in Section 4.12, this is an emergent property. To achieve it, an architect must select a specific \textit{constellation} of functional capabilities. Based on our taxonomy, this would likely involve the synergistic combination of:}
\begin{itemize}
    \item \textcolor{black}{A \texttt{Multi-path Plan Generator} to handle unforeseen obstacles.}
    \item \textcolor{black}{\texttt{Fully Distributed Coordination} to eliminate single points of failure.}
    \item \textcolor{black}{Advanced \texttt{Self-learning} capabilities to adapt without human intervention.}
\end{itemize}
\textcolor{black}{Using our decision matrix, an architect can then analyze the cumulative non-functional consequences of this ``high-autonomy'' architectural configuration. On one hand, this combination significantly enhances \textbf{Reliability} (\texttt{++}) by removing central points of failure and boosts the adaptability aspect of \textbf{Usability} (\texttt{++}) in dynamic environments. Conversely, these benefits are counterbalanced by significant drawbacks. The communication overhead inherent in \texttt{Fully Distributed Coordination} severely degrades \textbf{Performance} (\texttt{--}). The system's resultant complexity and non-determinism present considerable challenges for debugging, thus drastically reducing \textbf{Maintainability} (\texttt{--}). Furthermore, ensuring such a highly autonomous system behaves predictably and ethically introduces significant challenges for \textbf{Responsible \& Trustworthy AI}.}

\subsection{Illustrative Case Study: AI Audit Analyst}
\begin{figure*}[htbp]
    \centering
    \includegraphics[width=\textwidth]{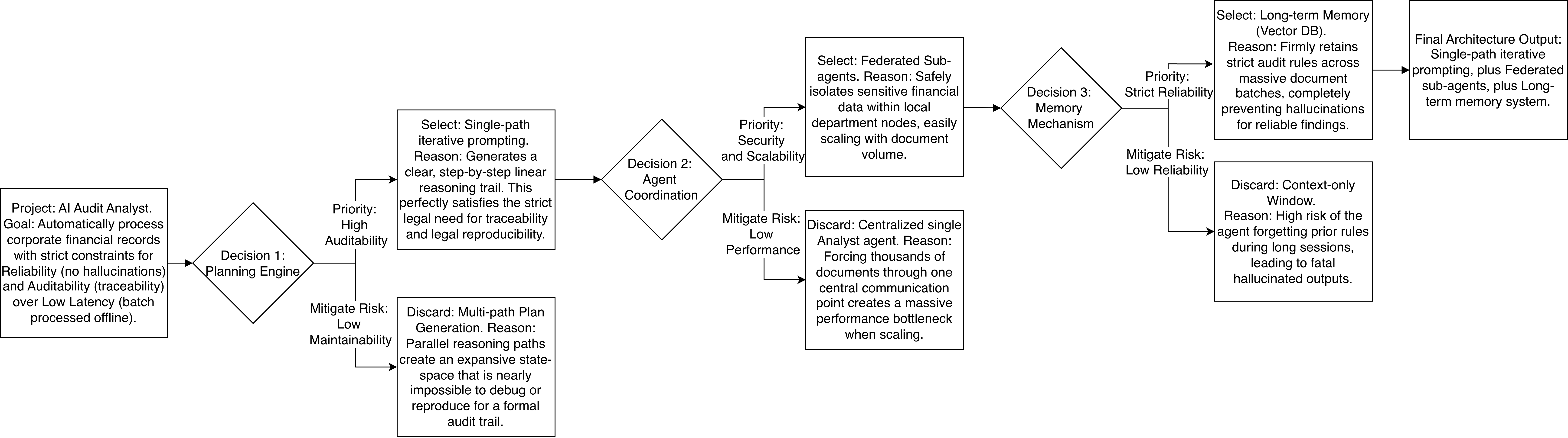}
    \caption{\textcolor{black}{Decision-making workflow for the AI Audit Analyst case study. The flowchart demonstrates the step-by-step traversal from initial non-functional requirements (NFRs) to the final composite architecture, illustrating how the decision matrix systematically resolves architectural conflicts across Planning, Coordination, and Memory dimensions.}}
    \label{fig:case_study_workflow}
\end{figure*}

\textcolor{black}{To demonstrate the practical utility and decision-making lifecycle of our model, we present a case study focused on developing an AI Audit Analyst, a system designed to autonomously process large volumes of corporate financial records to ensure regulatory compliance (Figure \ref{fig:case_study_workflow}). In this engineering context, the architect identifies two high-priority non-functional requirements: Reliability (the system must not hallucinate audit findings) and Auditability (every decision must be traceable for legal reasons). Conversely, the system has a lower priority for Latency, as audits are typically batch-processed offline.}

\textcolor{black}{Guided by the decision-making workflow, the architect first consults the taxonomy to identify candidate components for the Planning and Coordination dimensions. When evaluating Planning options, the architect faces a trade-off between Single-path and Multi-path reasoning. While Multi-path generators can improve task success rates, our Decision Matrix indicates a significant negative impact on Maintainability. This rating is supported by SLR evidence highlighting that parallel reasoning paths create an expansive state-space that is difficult to debug or reproduce during an audit trail. Prioritizing Auditability, the architect selects a Single-path iterative prompting mechanism with self-verification.}

\textcolor{black}{Next, for Agent Coordination, the architect evaluates Centralized versus Federated models. Although a Centralized model is easier to implement, the matrix reveals potential Performance bottlenecks when scaling to thousands of documents. By choosing a Federated approach, the architect achieves better Scalability and Security by isolating sensitive data within sub-agents, as suggested by practitioners in our validation survey.} 

\textcolor{black}{Finally, to guarantee the strict Reliability constraint over extended audit sessions, the architect evaluates the Memory dimension. While a Context-only approach is simpler, the decision matrix flags a severe risk to Reliability, as the agent might forget prior audit rules when processing large document batches. To prevent hallucinations and ensure the consistent application of standards, the architect selects Long-term Memory (e.g., via a Vector Database), accepting the trade-off of increased architectural Complexity. The resulting composite architecture—Single-path reasoning, Federated coordination, and Long-term memory—demonstrates how this structured traversal—from requirement prioritization to matrix-guided selection—transforms the theoretical framework into a practical handbook for resolving complex architectural conflicts.}

\section{Validation}
\label{sec:validation}
A multi-faceted approach was employed to validate the robustness, completeness, and practical applicability of the proposed taxonomy and decision model. \textcolor{black}{The validation relies primarily on a formal survey with a panel of experts, while its structural integrity is supported by the rigorous, evidence-based methodology used for its construction.}

\subsection{Methodological Groundedness and Structural Integrity}
\textcolor{black}{The structural integrity of our taxonomy is established through its rigorous construction process.} As detailed in Section 3, the entire study was guided by the systematic protocols for SLRs and the established taxonomy development guidelines from Usman et al.~\cite{usman2017taxonomies}, ensuring an empirically grounded foundation.

A primary outcome of this rigorous process was ensuring the taxonomy's \textbf{orthogonality}—the logical independence of its dimensions. This was a deliberate design goal, enforced by adhering to the principles of the Usman et al. methodology~\cite{usman2017taxonomies} (specifically steps B8-B11, as discussed in Section 3). This structured approach required a deliberate analysis of category relationships to ensure they were mutually exclusive. The success of this design choice was then formally confirmed through our external expert survey, as detailed in the next section.

\subsection{\textcolor{black}{External Validation: Expert-Driven Survey}}
\label{ssec:expert_survey}
To assess the taxonomy's quality and its practical relevance from an external perspective, we conducted a formal validation survey with a panel of experts from both industry and academia. We invited 12 experts to participate in the evaluation. Participants were identified through two primary channels: (1) academic experts selected for their scholarly contributions to the field, including authors of some primary studies in our SLR, and \textcolor{black}{(2) senior industry practitioners contacted via professional networks (including LinkedIn and specialized AI engineering communities), selected for their recognized, practical experience in designing and developing complex AI-based systems.} The panel comprised 6 academic researchers (including 2 full professors) in Software Engineering and AI, and 6 senior industry practitioners (including 4 software architects and 2 AI/ML engineers) with an average of 5+ years of experience in designing and developing complex AI-based systems. \textcolor{black}{The diverse backgrounds and high level of seniority within the panel ensure that the validation results are representative of both state-of-the-art research and industrial best practices. To ensure methodological rigor, the survey followed standard ethical guidelines: all participants provided informed consent, and their data were fully anonymized. For the analysis of open-ended feedback, we employed thematic analysis, where two researchers independently coded the responses and resolved any discrepancies through a negotiated agreement process. The profiles of these experts are summarized in Table~\ref{tab:expert_profiles}.}

\begin{table}[htbp]
\centering
\caption{\textcolor{black}{Profiles of Experts in the External Validation Survey}}
\label{tab:expert_profiles}
\begin{tabularx}{\columnwidth}{@{}lXclc@{}}
\toprule
\textbf{Affiliation} & \textbf{Professional Role} & \textbf{Year} & \textbf{Expertise Level} & \textbf{Count} \\ \midrule
Academia             & Full Professor             & 15+           & Senior/Expert            & 2              \\
                     & Senior Researcher          & 5+            & Mid-Senior               & 4              \\ \midrule
Industry             & Senior Software Architect  & 10+           & Senior/Expert            & 4              \\
                     & AI/ML Engineer             & 5+            & Professional             & 2              \\ \midrule
\textbf{Total}       &                            &               &                          & \textbf{12}    \\ \bottomrule
\end{tabularx}
\end{table}

We designed a structured questionnaire (The questionnaire is included in the appendix. The full instrument is available in our replication package) to guide the evaluation. The experts were provided with our manuscript, the main taxonomy figures, and the decision matrix. The questionnaire was divided into two main parts. For the taxonomy, we asked experts to evaluate its completeness, correctness, and orthogonality; for the decision model, we focused on its clarity and practical utility. Feedback was collected through a mix of five-point Likert scale ratings (from 1=Strongly Disagree to 5=Strongly Agree) and open-ended questions.

\subsubsection{\textcolor{black}{Results of the Taxonomy Evaluation}}
\textcolor{black}{The first part of the survey assessed the taxonomy based on three key criteria, and the quantitative results indicate a strong positive consensus from the expert panel. The full survey instrument and detailed results are available in our online replication package.}
\begin{itemize}
    \item \textbf{Completeness:} \textcolor{black}{This criterion evaluated whether the taxonomy covers all critical architectural aspects of FM-based agents. The mean score from the 12 experts was 4.63 (out of 5.0), with a standard deviation of 0.53, indicating strong agreement.}
    \item \textbf{Correctness:} \textcolor{black}{This assessed whether the dimensions and categories are logically sound and well-defined. The mean score was 4.59, with a standard deviation of 0.66, reflecting a very high level of agreement on the taxonomy's logical integrity.}
    \item \textbf{Orthogonality:} \textcolor{black}{While experts affirmed the clear independence of foundational categories such as Input Modality, their comments on more interconnected areas were particularly elucidating. For instance, the discussion around the procedural connection between Planning and Action, and the conceptual proximity of Reliability, Robustness, and Resilience, underscored the value of our taxonomy's deliberate distinctions. Our taxonomy intentionally separates these design options to provide architects with the necessary precision to distinguish high-level strategy from runtime execution. Thus, the feedback confirmed that our approach to orthogonality is not based on creating isolated silos, but on providing the clear and essential distinctions needed for practical architectural reasoning.}

\end{itemize}

\subsubsection{Results of the Decision Model Evaluation}
\textcolor{black}{The second part of the survey focused on the Decision Matrix, which was also received very positively by the experts, particularly for its practical applicability.}
\begin{itemize}
    \item \textbf{Clarity:} \textcolor{black}{This criterion evaluated whether the model clearly illustrates the architectural trade-offs. The mean score was 4.17, with a standard deviation of 0.72, indicating exceptional agreement on its clarity.}
    \item \textbf{Utility:} \textcolor{black}{This assessed whether the model is perceived as a practical and useful tool for architects. The mean score was 4.58, with a standard deviation of 0.67, confirming its high practical value.}
\end{itemize}

\textcolor{black}{The qualitative feedback strongly supported the Decision Model's value as a practical tool for architectural reasoning. Experts praised the model for making critical trade-offs "explicit", a feature one participant noted is "often missing in academic frameworks". This clarity was attributed to the matrix format, which makes complex architectural compromises "visible at a glance," thereby enabling more informed and deliberate design choices. The constructive suggestions for future enhancements further affirmed the model's potential. These included exploring ways to quantify the impacts of design choices, extending the model with a \texttt{Cost} dimension, and providing common mitigation patterns for negative trade-offs. This feedback confirms the model's role as a practical and extensible foundation for future architectural research.}

\section{Threats to Validity}
\label{sec:threats}
\textcolor{black}{We discuss the potential threats to the validity of our study using the well-established four-category framework for software engineering research \cite{lago2024threats}. We detail each potential threat and the mitigation strategies we employed.}

\subsection{Construct Validity}
\textcolor{black}{Construct validity concerns the extent to which our study measures what it claims to be measuring. A primary threat here is that our search keywords might not have captured all relevant studies on FM-based agent architectures. To mitigate this, we employed a multi-pronged search strategy. We used the Quasi-Gold Standard (QGS) method to systematically derive our search string from a known set of highly relevant papers. Furthermore, we supplemented our keyword search with both forward and backward snowballing to identify relevant studies that may have been missed.}

\subsection{Internal Validity}
\textcolor{black}{Internal validity relates to confounding factors that could influence the study's results, primarily researcher bias. A threat exists in the subjective nature of study selection and data extraction. To mitigate this, we designed a rigorous protocol with several safeguards. We used a clear and objective set of inclusion and exclusion criteria (Table 1). The classification and data extraction process involved an initial classification by one researcher, followed by a complete review and verification by a second, senior researcher. Any disagreements were resolved through consensus-based discussion among all authors to ensure consistency.}

\textcolor{black}{A potential threat to internal validity relates to the qualitative coding process used to develop the taxonomy. As described in Section 3.1.5, we employed a negotiated agreement approach rather than independent double-coding. Consequently, we did not calculate an inter-rater reliability statistic such as Cohen's Kappa. While our consensus-driven approach facilitated deep discussions and resolution of complex architectural nuances, we acknowledge that the lack of independent double-coding represents a limitation in demonstrating strict quantitative reliability of the coding process.} \textcolor{black}{This limitation is not confined to taxonomy coding: study classification and data extraction, and the quality assessment scoring in Section 3.2.1, followed the same one-researcher-then-audit process rather than independent double-coding. We acknowledge this as a methodological limitation across our subjective analysis tasks and recommend that future replications incorporate independent double-coding with reliability metrics throughout the pipeline.}

The impact ratings in the Decision Matrix are derived from a qualitative synthesis of literature and researcher observations. This introduces a potential risk of subjective bias in how trade-offs were interpreted and mapped. We mitigated this risk by implementing a dual-researcher cross-check protocol and grounding the ratings in the "Non-functional Qualities" evidence recorded during data extraction. Furthermore, the expert survey served as an external sanity check; however, we acknowledge that the survey focused on high-level consensus rather than a granular validation of every individual cell in the matrix.

\textcolor{black}{We acknowledge that the exclusion of papers by GPT-4o was not independently supervised at the time. The use of GPT-4o during screening may have introduced false exclusions or systematic bias. Although the primary researcher reviewed the filtering results at each round and retained responsibility for all final decisions, we did not conduct a separate sampling audit of the excluded papers to estimate agreement with an independent manual assessment. We acknowledge this as a limitation and recommend that future reviews using LLM-assisted screening include a documented random audit of excluded studies.}

\subsection{External Validity}
\textcolor{black}{External validity concerns the generalizability of our findings. We identify four primary threats here:}
\begin{itemize}
    \item \textcolor{black}{\textbf{Reliance on Non-Peer-Reviewed Pre-prints:} This is a major threat, as a significant portion of our primary studies are pre-prints from sources like arXiv, which do not undergo formal peer review. This was a deliberate methodological choice. In a fast-evolving field like FM-based agents, strictly limiting our review to peer-reviewed literature would have resulted in an outdated and less relevant taxonomy. To mitigate the risk of including low-quality work, we implemented a stringent, differentiated \textbf{quality assessment (QA) protocol} (detailed in Section 3.2.1), which ensured all included studies, regardless of publication status, met a high and consistent standard of research quality.}

    \textcolor{black}{A potential limitation of our differentiated scoring approach is that QAQ9 was assessed only for peer-reviewed studies. Consequently, a peer-reviewed paper published in a less established venue could receive a lower QAQ9 score, whereas this venue-related dimension was excluded from the normalised score of a pre-print. Although this design avoids treating an incomplete publication process as evidence of low methodological quality, it may introduce some imbalance between the two assessment schemes. We mitigated this concern by applying comparable proportional thresholds to the quality dimensions applicable to each publication type and by clearly distinguishing pre-prints from peer-reviewed studies throughout the analysis. In practice, this means a pre-print is not penalised on this dimension in the way a peer-reviewed paper from a less established venue is, effectively placing it on par with a peer-reviewed paper from a reputable venue on this specific dimension.}
    
    \item \textcolor{black}{\textbf{Exclusion of Grey Literature:} As our SLR focused predominantly on academic literature, it explicitly excluded grey literature such as GitHub repositories, Hugging Face model cards, and engineering blogs. Consequently, our taxonomy may better represent the architectural patterns of academic research agents rather than those deployed in real-world industrial settings. We acknowledge this as a limitation and identify the incorporation of these practical sources as a critical avenue for future work.}
    \item \textcolor{black}{\textbf{Limited Scope and Domain Coverage of the Expert Validation:} Our expert validation was conducted with a panel of 12 experts. Although the panel was selected to provide a balanced academic and industry perspective, their views may not represent the wider software architecture and AI community. The scores reported in Section 6.2 should therefore be interpreted as agreement within this panel, rather than as evidence that the taxonomy and Decision Matrix apply equally well across all application domains. Highly specialised domains, such as safety-critical embedded systems and medical devices, are underrepresented both in our 66 primary studies and in the backgrounds of the panel. The applicability of the Decision Matrix may therefore vary in contexts with project constraints or technologies not fully represented in our evidence base. We also did not conduct other forms of validation, such as surveying the authors of the primary studies. We acknowledge these limitations and identify broader and domain-stratified validation as directions for future work.}
    \item \textcolor{black}{\textbf{Temporal Validity and Taxonomy Evolution:} The rapid pace of FM advancements poses a threat to the long-term relevance of any static taxonomy. To mitigate this, we designed our taxonomy with a hierarchical structure that ensures extensibility. This allows new architectural patterns or emerging capabilities (e.g., multi-modal reasoning) to be integrated as new leaf nodes without restructuring the core dimensions. Furthermore, we acknowledge that the architectural trade-offs captured in our Decision Model are subject to contextual constraints; the impact of specific components on system latency or reliability may vary depending on the underlying model's efficiency and the specific application domain, ensuring the framework remains adaptable as the field evolves.}
\end{itemize}

\subsection{Conclusion Validity}
Conclusion validity concerns the degree to which conclusions about relationships in our data are reasonable. A threat here is the potential for subjectivity in the thematic analysis and synthesis process used to create the taxonomy from the extracted data. To mitigate this, the terminology control and the grouping of concepts into categories and dimensions were not performed by a single researcher. It was a collaborative process involving multiple rounds of discussion and consensus-building among the authors to ensure the resulting taxonomy structure was a logical and defensible synthesis of the evidence from the literature. \textcolor{black}{This threat also extends to our Decision Model. The model is intended as a high-level reasoning tool for making architectural trade-offs explicit, rather than as a precise quantitative predictor. Synthesising qualitative evidence from multiple studies into impact ratings necessarily involves interpretation and may introduce researcher bias. The model is also constrained by the scope and limitations of the primary literature from which it was derived. Accordingly, the ratings should be interpreted as generalised baselines. Where the primary studies reported mixed results for an architectural option, the assigned rating reflects the most prevalent trend across the dataset, as described in Section 5.1. It therefore indicates the likely direction of a trade-off rather than its magnitude in a specific system. In such borderline cases, architects should use the rating as a starting point for further analysis against their own project constraints, rather than as a definitive answer. Quantifying these impacts, as also suggested by participants in our validation survey, remains an important direction for future work.}

\section{Discussion}
\label{sec:discussion}
In this section, we discuss the potential implications of our taxonomy and decision model for three key stakeholder groups: academic researchers, software developers and architects, and foundation model (FM) vendors and tool builders. \textcolor{black}{Through these perspectives, we explicitly revisit and synthesize our findings for RQ1 (Architectural Elements), RQ2 (Trade-offs), and RQ3 (Design Guidance).}

\subsection{Implications for Researchers}
For academic researchers, our work offers several benefits. First, the \textbf{taxonomy provides a common vocabulary and a structured intellectual framework} for the community. It can be used to classify new agent architectures, enabling clearer comparisons and facilitating a more systematic accumulation of knowledge. Second, by mapping the current landscape, our work helps to \textbf{identify under-explored areas and potential research gaps}. For instance, certain combinations of architectural options may be rare in the literature, suggesting avenues for novel research. Finally, our \textbf{Decision Model serves as a set of falsifiable hypotheses} about the relationships between architectural choices and system qualities, which can be validated, refined, or challenged in future empirical studies. \textcolor{black}{Specifically, researchers can use the taxonomy dimensions as a basis for defining 'control groups' when benchmarking the performance of different agent architectures.}

\textcolor{black}{Takeaway 1 (RQ1): Our taxonomy provides a multi-dimensional framework that formalizes the design space of FM-based agents. It transitions the field from ad-hoc prompting to a systematic "Architectural Engineering" paradigm, offering a common vocabulary for future empirical research.}

\subsection{Implications for Developers and Architects}
For practitioners responsible for building agent-based systems, our contributions serve as a practical guide. The \textbf{taxonomy acts as a design space map or a "pattern catalogue,"} presenting architects with a clear and comprehensive set of available architectural options. This can prevent ad-hoc design and encourage a more structured design process. More importantly, the \textbf{Decision Model is a practical tool for architectural reasoning}. It forces a consideration of non-functional qualities early in the design process and makes the inherent trade-offs explicit (e.g., choosing a `Multi-path Plan Generator` for higher adaptability may decrease `Maintainability`). This allows development teams to make more informed, deliberate design decisions that are consciously aligned with their specific system requirements. \textcolor{black}{In practice, architects can utilize the Decision Model as a 'pre-flight checklist' during design reviews to systematically evaluate whether the selected components (e.g., Memory, Planning) meet the project’s specific latency and reliability constraints.}

\textcolor{black}{Takeaway 2 (RQ2): Architectural decisions are characterized by inherent trade-offs (e.g., Autonomy vs. Maintainability). Our Decision Model provides a structured way to navigate these compromises based on specific system-level quality requirements.}

\subsection{Implications for FM Vendors and Tool Builders}
Our work also has direct implications for the vendors who create foundational models and the builders of popular agent frameworks. The taxonomy provides them with a \textbf{structured view of how their technologies are being used} in real-world agent architectures. By seeing which architectural options are most frequently combined, they can identify common use-case patterns and prioritize the development of features, APIs, or components that better support these patterns. For example, understanding the trade-offs in different `Memory` options can guide a tool builder in designing a more versatile and efficient memory module. Ultimately, our framework can help align the evolution of the underlying platforms and tools with the practical needs of architects and developers. \textcolor{black}{For instance, tool builders can use the identified 'rare combinations' in our taxonomy to discover underserved architectural patterns, guiding the development of new standardized APIs or middleware that simplify the integration of complex reasoning loops.}

\textcolor{black}{Takeaway 3 (RQ3): The transition from ad-hoc integration to structured design is facilitated by our framework, which acts as a practical guidance for building next-generation, standardized AI agent platforms.}

\section{Conclusion}
\label{sec:con}

\textcolor{black}{FM-based agents are gaining increasing attention in various domains. Our study reveals that the development of these agents is transitioning from ad-hoc experimentation to a structured architectural discipline. By formalizing this space, our taxonomy and decision model extend fundamental software engineering principles—specifically the "Separation of Concerns"—to the stochastic nature of AI-native systems. This taxonomy provides a shared vocabulary that not only addresses standardization issues but also offers a theoretical bridge for applying established design rigor to foundation-model-based architectures. To ensure its ongoing relevance, the taxonomy is designed as an extensible 'living framework', capable of evolving alongside advancements in foundation models.} 

In future work, we will explore how to apply this taxonomy in conjunction with existing architectural patterns. This study, anchored in a systematic literature review and expert validation, provides a foundational framework. However, as the field of FM-based agents is rapidly evolving, we see significant value in extending and further validating this research through a wider range of empirical methods. Future studies could involve more practitioner-oriented approaches, such as in-depth interviews with industry architects to refine the taxonomy's practical relevance and large-scale mining studies of open-source agent repositories. Further validation could also be achieved by surveying the authors of our primary studies for their direct feedback and by replicating our SLR process with a specific focus on high-quality grey literature to broaden the evidence base. \textcolor{black}{Given the limited size and domain coverage of our validation panel, future work should conduct domain-stratified validation with practitioners from distinct and highly specialised application domains to assess how well the taxonomy and Decision Matrix generalise beyond the contexts covered in this study.} We believe such future work will build upon the foundational classification we have established and further bridge the gap between architectural theory and practice.

\textcolor{black}{\textbf{Declaration of generative AI and AI-assisted technologies in the manuscript preparation process:} During the preparation of this work the author(s) used GPT-4o in order to assist the SLR process. After using this tool/service, the author(s) reviewed and edited the content as needed and take(s) full responsibility for the content of the published article.}

\bibliographystyle{elsarticle-num}
\bibliography{reference}

@article{hong2023metagpt,
  title={Metagpt: Meta programming for multi-agent collaborative framework},
  author={Hong, Sirui and Zheng, Xiawu and Chen, Jonathan and Cheng, Yuheng and Wang, Jinlin and Zhang, Ceyao and Wang, Zili and Yau, Steven Ka Shing and Lin, Zijuan and Zhou, Liyang and others},
  journal={arXiv preprint arXiv:2308.00352},
  year={2023}
}

@article{liu2023agentbench,
  title={Agentbench: Evaluating llms as agents},
  author={Liu, Xiao and Yu, Hao and Zhang, Hanchen and Xu, Yifan and Lei, Xuanyu and Lai, Hanyu and Gu, Yu and Ding, Hangliang and Men, Kaiwen and Yang, Kejuan and others},
  journal={arXiv preprint arXiv:2308.03688},
  year={2023}
}

@manual{iso25010,
  title        = {{ISO/IEC 25010:2023 Systems and Software Engineering --- Systems and
                  Software Quality Requirements and Evaluation (SQuaRE) --- Product
                  Quality Model}},
  organization = {International Organization for Standardization},
  address      = {Geneva, Switzerland},
  edition      = {2},
  year         = {2023},
  howpublished = {\url{https://www.iso.org/standard/78176.html}}
}

@book{bass2021architecture,
  author    = {Bass, Len and Clements, Paul and Kazman, Rick},
  title     = {Software Architecture in Practice},
  edition   = {4},
  publisher = {Addison-Wesley Professional},
  address   = {Boston, MA},
  year      = {2021}
}

@book{bass2025engineering,
  author    = {Bass, Len and Lu, Qinghua and Weber, Ingo and Zhu, Liming},
  title     = {Engineering AI Systems: Architecture and DevOps Essentials},
  publisher = {Addison-Wesley Professional},
  address   = {Boston, MA},
  year      = {2025}
}

@misc{gemini2026,
  author       = {{Google DeepMind}},
  title        = {Gemini},
  year         = {2026},
  note         = {Accessed: 2026-07-30},
  howpublished = {\url{https://deepmind.google/models/gemini/}}
}

@misc{llama4,
  author       = {{Meta AI}},
  title        = {The Llama 4 Herd: The Beginning of a New Era of Natively Multimodal AI Innovation},
  year         = {2025},
  note         = {Accessed: 2026-07-30},
  howpublished = {\url{https://ai.meta.com/blog/llama-4-multimodal-intelligence/}}
}

@inproceedings{chen2023agentverse,
  title={Agentverse: Facilitating multi-agent collaboration and exploring emergent behaviors},
  author={Chen, Weize and Su, Yusheng and Zuo, Jingwei and Yang, Cheng and Yuan, Chenfei and Chan, Chi-Min and Yu, Heyang and Lu, Yaxi and Hung, Yi-Hsin and Qian, Chen and others},
  booktitle={The Twelfth International Conference on Learning Representations},
  year={2023}
}

@article{vaswani2017attention,
  title={Attention is all you need},
  author={Vaswani, Ashish and Shazeer, Noam and Parmar, Niki and Uszkoreit, Jakob and Jones, Llion and Gomez, Aidan N and Kaiser, {\L}ukasz and Polosukhin, Illia},
  journal={Advances in neural information processing systems},
  volume={30},
  year={2017}
}

@article{hou2024large,
  title={Large language models for software engineering: A systematic literature review},
  author={Hou, Xinyi and Zhao, Yanjie and Liu, Yue and Yang, Zhou and Wang, Kailong and Li, Li and Luo, Xiapu and Lo, David and Grundy, John and Wang, Haoyu},
  journal={ACM Transactions on Software Engineering and Methodology},
  volume={33},
  number={8},
  pages={1--79},
  year={2024},
  publisher={ACM New York, NY}
}

@article{wang2022machine,
  title={Machine/deep learning for software engineering: A systematic literature review},
  author={Wang, Simin and Huang, Liguo and Gao, Amiao and Ge, Jidong and Zhang, Tengfei and Feng, Haitao and Satyarth, Ishna and Li, Ming and Zhang, He and Ng, Vincent},
  journal={IEEE Transactions on Software Engineering},
  volume={49},
  number={3},
  pages={1188--1231},
  year={2022},
  publisher={IEEE}
}

@article{naveed2024model,
  title={Model driven engineering for machine learning components: A systematic literature review},
  author={Naveed, Hira and Arora, Chetan and Khalajzadeh, Hourieh and Grundy, John and Haggag, Omar},
  journal={Information and Software Technology},
  volume={169},
  pages={107423},
  year={2024},
  publisher={Elsevier}
}

@article{ZHANG2011625,
title = {Identifying relevant studies in software engineering},
journal = {Information and Software Technology},
volume = {53},
number = {6},
pages = {625-637},
year = {2011},
note = {Special Section: Best papers from the APSEC},
issn = {0950-5849},
doi = {https://doi.org/10.1016/j.infsof.2010.12.010},
url = {https://www.sciencedirect.com/science/article/pii/S0950584910002260},
author = {He Zhang and Muhammad Ali Babar and Paolo Tell}
}

@article{usman2017taxonomies,
  title={Taxonomies in software engineering: A systematic mapping study and a revised taxonomy development method},
  author={Usman, Muhammad and Britto, Ricardo and B{\"o}rstler, J{\"u}rgen and Mendes, Emilia},
  journal={Information and Software Technology},
  volume={85},
  pages={43--59},
  year={2017},
  publisher={Elsevier}
}

@article{wu2024copilot,
  title={Os-copilot: Towards generalist computer agents with self-improvement},
  author={Wu, Zhiyong and Han, Chengcheng and Ding, Zichen and Weng, Zhenmin and Liu, Zhoumianze and Yao, Shunyu and Yu, Tao and Kong, Lingpeng},
  journal={arXiv preprint arXiv:2402.07456},
  year={2024}
}

@article{wu2024evolutionary,
  title={Evolutionary Computation in the Era of Large Language Model: Survey and Roadmap},
  author={Wu, Xingyu and Wu, Sheng-hao and Wu, Jibin and Feng, Liang and Tan, Kay Chen},
  journal={arXiv preprint arXiv:2401.10034},
  year={2024}
}

@article{lu2023routing,
  title={Routing to the expert: Efficient reward-guided ensemble of large language models},
  author={Lu, Keming and Yuan, Hongyi and Lin, Runji and Lin, Junyang and Yuan, Zheng and Zhou, Chang and Zhou, Jingren},
  journal={arXiv preprint arXiv:2311.08692},
  year={2023}
}

@article{gao2024memory,
  title={Memory Sharing for Large Language Model based Agents},
  author={Gao, Hang and Zhang, Yongfeng},
  journal={arXiv preprint arXiv:2404.09982},
  year={2024}
}

@article{maharana2024evaluating,
  title={Evaluating very long-term conversational memory of llm agents},
  author={Maharana, Adyasha and Lee, Dong-Ho and Tulyakov, Sergey and Bansal, Mohit and Barbieri, Francesco and Fang, Yuwei},
  journal={arXiv preprint arXiv:2402.17753},
  year={2024}
}

@article{xi2023rise,
  title={The rise and potential of large language model based agents: A survey},
  author={Xi, Zhiheng and Chen, Wenxiang and Guo, Xin and He, Wei and Ding, Yiwen and Hong, Boyang and Zhang, Ming and Wang, Junzhe and Jin, Senjie and Zhou, Enyu and others},
  journal={arXiv preprint arXiv:2309.07864},
  year={2023}
}

@article{baechler2024screenai,
  title={Screenai: A vision-language model for ui and infographics understanding},
  author={Baechler, Gilles and Sunkara, Srinivas and Wang, Maria and Zubach, Fedir and Mansoor, Hassan and Etter, Vincent and C{\u{a}}rbune, Victor and Lin, Jason and Chen, Jindong and Sharma, Abhanshu},
  journal={arXiv preprint arXiv:2402.04615},
  year={2024}
}

@article{zhang2024survey,
  title={A survey on the memory mechanism of large language model based agents},
  author={Zhang, Zeyu and Bo, Xiaohe and Ma, Chen and Li, Rui and Chen, Xu and Dai, Quanyu and Zhu, Jieming and Dong, Zhenhua and Wen, Ji-Rong},
  journal={arXiv preprint arXiv:2404.13501},
  year={2024}
}

@article{guo2024large,
  title={Large language model based multi-agents: A survey of progress and challenges},
  author={Guo, Taicheng and Chen, Xiuying and Wang, Yaqi and Chang, Ruidi and Pei, Shichao and Chawla, Nitesh V and Wiest, Olaf and Zhang, Xiangliang},
  journal={arXiv preprint arXiv:2402.01680},
  year={2024}
}

@article{wu2023bloomberggpt,
  title={Bloomberggpt: A large language model for finance},
  author={Wu, Shijie and Irsoy, Ozan and Lu, Steven and Dabravolski, Vadim and Dredze, Mark and Gehrmann, Sebastian and Kambadur, Prabhanjan and Rosenberg, David and Mann, Gideon},
  journal={arXiv preprint arXiv:2303.17564},
  year={2023}
}

@article{du2024survey,
  title={A Survey on Context-Aware Multi-Agent Systems: Techniques, Challenges and Future Directions},
  author={Du, Hung and Thudumu, Srikanth and Vasa, Rajesh and Mouzakis, Kon},
  journal={arXiv preprint arXiv:2402.01968},
  year={2024}
}

@article{zhang2024codeagent,
  title={Codeagent: Enhancing code generation with tool-integrated agent systems for real-world repo-level coding challenges},
  author={Zhang, Kechi and Li, Jia and Li, Ge and Shi, Xianjie and Jin, Zhi},
  journal={arXiv preprint arXiv:2401.07339},
  year={2024}
}

@article{jimenez2023swe,
  title={Swe-bench: Can language models resolve real-world github issues?},
  author={Jimenez, Carlos E and Yang, John and Wettig, Alexander and Yao, Shunyu and Pei, Kexin and Press, Ofir and Narasimhan, Karthik},
  journal={arXiv preprint arXiv:2310.06770},
  year={2023}
}

@inproceedings{durante2024interactive,
  title={An interactive agent foundation model},
  author={Durante, Zane and Gong, Ran and Sarkar, Bidipta and Wake, Naoki and Taori, Rohan and Tang, Paul and Lakshmikanth, Shrinidhi and Schulman, Kevin and Milstein, Arnold and Vo, Hoi and others},
  booktitle={Proceedings of the Computer Vision and Pattern Recognition Conference},
  pages={3652--3662},
  year={2025}
}

@article{li2024survey,
  title={A Survey on LLM-Based Agents: Common Workflows and Reusable LLM-Profiled Components},
  author={Li, Xinzhe},
  journal={arXiv preprint arXiv:2406.05804},
  year={2024}
}

@article{hao2023reasoning,
  title={Reasoning with language model is planning with world model},
  author={Hao, Shibo and Gu, Yi and Ma, Haodi and Hong, Joshua Jiahua and Wang, Zhen and Wang, Daisy Zhe and Hu, Zhiting},
  journal={arXiv preprint arXiv:2305.14992},
  year={2023}
}

@article{yang2023foundation,
  title={Foundation models for decision making: Problems, methods, and opportunities},
  author={Yang, Sherry and Nachum, Ofir and Du, Yilun and Wei, Jason and Abbeel, Pieter and Schuurmans, Dale},
  journal={arXiv preprint arXiv:2303.04129},
  year={2023}
}

@inproceedings{lu2024towards,
  title={Towards responsible generative ai: A reference architecture for designing foundation model based agents},
  author={Lu, Qinghua and Zhu, Liming and Xu, Xiwei and Xing, Zhenchang and Harrer, Stefan and Whittle, Jon},
  booktitle={2024 IEEE 21st International Conference on Software Architecture Companion (ICSA-C)},
  pages={119--126},
  year={2024},
  organization={IEEE}
}

@inproceedings{brand2023towards,
  title={Towards Improved User Experience for Artificial Intelligence Systems},
  author={Brand, Lisa and Humm, Bernhard G and Krajewski, Andrea and Zender, Alexander},
  booktitle={International Conference on Engineering Applications of Neural Networks},
  pages={33--44},
  year={2023},
  organization={Springer}
}

@inproceedings{yang2024generalized,
  title={Generalized predictive model for autonomous driving},
  author={Yang, Jiazhi and Gao, Shenyuan and Qiu, Yihang and Chen, Li and Li, Tianyu and Dai, Bo and Chitta, Kashyap and Wu, Penghao and Zeng, Jia and Luo, Ping and others},
  booktitle={Proceedings of the IEEE/CVF Conference on Computer Vision and Pattern Recognition},
  pages={14662--14672},
  year={2024}
}

@article{liu2024position,
  title={Position: Foundation Agents as the Paradigm Shift for Decision Making},
  author={Liu, Xiaoqian and Lou, Xingzhou and Jiao, Jianbin and Zhang, Junge},
  journal={arXiv preprint arXiv:2405.17009},
  year={2024}
}

@article{qian2023communicative,
  title={Communicative agents for software development},
  author={Qian, Chen and Cong, Xin and Yang, Cheng and Chen, Weize and Su, Yusheng and Xu, Juyuan and Liu, Zhiyuan and Sun, Maosong},
  journal={arXiv preprint arXiv:2307.07924},
  year={2023}
}

@article{yao2024tree,
  title={Tree of thoughts: Deliberate problem solving with large language models},
  author={Yao, Shunyu and Yu, Dian and Zhao, Jeffrey and Shafran, Izhak and Griffiths, Tom and Cao, Yuan and Narasimhan, Karthik},
  journal={Advances in Neural Information Processing Systems},
  volume={36},
  year={2024}
}

@article{modarressi2023ret,
  title={Ret-llm: Towards a general read-write memory for large language models},
  author={Modarressi, Ali and Imani, Ayyoob and Fayyaz, Mohsen and Sch{\"u}tze, Hinrich},
  journal={arXiv preprint arXiv:2305.14322},
  year={2023}
}

@article{islam2024mapcoder,
  title={Mapcoder: Multi-agent code generation for competitive problem solving},
  author={Islam, Md Ashraful and Ali, Mohammed Eunus and Parvez, Md Rizwan},
  journal={arXiv preprint arXiv:2405.11403},
  year={2024}
}

@article{huang2025comprehensive,
  title={Comprehensive Fine-Tuning Large Language Models of Code for Automated Program Repair},
  author={Huang, Kai and Zhang, Jian and Bao, Xinlei and Wang, Xu and Liu, Yang},
  journal={IEEE Transactions on Software Engineering},
  year={2025},
  publisher={IEEE}
}

@article{biagiola2024testing,
  title={Testing of deep reinforcement learning agents with surrogate models},
  author={Biagiola, Matteo and Tonella, Paolo},
  journal={ACM Transactions on Software Engineering and Methodology},
  volume={33},
  number={3},
  pages={1--33},
  year={2024},
  publisher={ACM New York, NY, USA}
}

@inproceedings{feng2024prompting,
  title={Prompting is all you need: Automated android bug replay with large language models},
  author={Feng, Sidong and Chen, Chunyang},
  booktitle={Proceedings of the 46th IEEE/ACM International Conference on Software Engineering},
  pages={1--13},
  year={2024}
}

@article{zhang2025empowering,
  title={Empowering agile-based generative software development through human-ai teamwork},
  author={Zhang, Sai and Xing, Zhenchang and Guo, Ronghui and Xu, Fangzhou and Chen, Lei and Zhang, Zhaoyuan and Zhang, Xiaowang and Feng, Zhiyong and Zhuang, Zhiqiang},
  journal={ACM Transactions on Software Engineering and Methodology},
  year={2025},
  publisher={ACM New York, NY}
}

@inproceedings{takagi2025framework,
  title={A framework for efficient development and debugging of role-playing agents with large language models},
  author={Takagi, Hirohane and Moriya, Shoji and Sato, Takuma and Nagao, Manabu and Higuchi, Keita},
  booktitle={Proceedings of the 30th International Conference on Intelligent User Interfaces},
  pages={70--88},
  year={2025}
}

@article{mostajabdaveh2024optimization,
  title={Optimization modeling and verification from problem specifications using a multi-agent multi-stage LLM framework},
  author={Mostajabdaveh, Mahdi and Yu, Timothy T and Ramamonjison, Rindranirina and Carenini, Giuseppe and Zhou, Zirui and Zhang, Yong},
  journal={INFOR: Information Systems and Operational Research},
  volume={62},
  number={4},
  pages={599--617},
  year={2024},
  publisher={Taylor \& Francis}
}

@inproceedings{chang2025drc,
  title={DRC-Coder: Automated drc checker code generation using LLM autonomous agent},
  author={Chang, Chen-Chia and Ho, Chia-Tung and Li, Yaguang and Chen, Yiran and Ren, Haoxing},
  booktitle={Proceedings of the 2025 International Symposium on Physical Design},
  pages={143--151},
  year={2025}
}

@article{zhu2024survey,
  title={A survey of multi-agent deep reinforcement learning with communication},
  author={Zhu, Changxi and Dastani, Mehdi and Wang, Shihan},
  journal={Autonomous Agents and Multi-Agent Systems},
  volume={38},
  number={1},
  pages={4},
  year={2024},
  publisher={Springer}
}

@inproceedings{zhong2024memorybank,
  title={Memorybank: Enhancing large language models with long-term memory},
  author={Zhong, Wanjun and Guo, Lianghong and Gao, Qiqi and Ye, He and Wang, Yanlin},
  booktitle={Proceedings of the AAAI Conference on Artificial Intelligence},
  volume={38},
  number={17},
  pages={19724--19731},
  year={2024}
}

@article{sumers2023cognitive,
  title={Cognitive architectures for language agents},
  author={Sumers, Theodore R and Yao, Shunyu and Narasimhan, Karthik and Griffiths, Thomas L},
  journal={arXiv preprint arXiv:2309.02427},
  year={2023}
}

@article{wang2023voyager,
  title={Voyager: An open-ended embodied agent with large language models},
  author={Wang, Guanzhi and Xie, Yuqi and Jiang, Yunfan and Mandlekar, Ajay and Xiao, Chaowei and Zhu, Yuke and Fan, Linxi and Anandkumar, Anima},
  journal={arXiv preprint arXiv:2305.16291},
  year={2023}
}

@article{yao2022react,
  title={React: Synergizing reasoning and acting in language models},
  author={Yao, Shunyu and Zhao, Jeffrey and Yu, Dian and Du, Nan and Shafran, Izhak and Narasimhan, Karthik and Cao, Yuan},
  journal={arXiv preprint arXiv:2210.03629},
  year={2022}
}

@article{guan2023leveraging,
  title={Leveraging pre-trained large language models to construct and utilize world models for model-based task planning},
  author={Guan, Lin and Valmeekam, Karthik and Sreedharan, Sarath and Kambhampati, Subbarao},
  journal={Advances in Neural Information Processing Systems},
  volume={36},
  pages={79081--79094},
  year={2023}
}

@article{pesce2023learning,
  title={Learning multi-agent coordination through connectivity-driven communication},
  author={Pesce, Emanuele and Montana, Giovanni},
  journal={Machine Learning},
  volume={112},
  number={2},
  pages={483--514},
  year={2023},
  publisher={Springer}
}

@article{agashe2023evaluating,
  title={Evaluating multi-agent coordination abilities in large language models},
  author={Agashe, Saaket and Fan, Yue and Wang, Xin Eric},
  journal={arXiv preprint arXiv:2310.03903},
  year={2023}
}

@article{gupta2024essential,
  title={The essential role of causality in foundation world models for embodied ai},
  author={Gupta, Tarun and Gong, Wenbo and Ma, Chao and Pawlowski, Nick and Hilmkil, Agrin and Scetbon, Meyer and Rigter, Marc and Famoti, Ade and Llorens, Ashley Juan and Gao, Jianfeng and others},
  journal={arXiv preprint arXiv:2402.06665},
  year={2024}
}

@inproceedings{you2025ferret,
  title={Ferret-ui: Grounded mobile ui understanding with multimodal llms},
  author={You, Keen and Zhang, Haotian and Schoop, Eldon and Weers, Floris and Swearngin, Amanda and Nichols, Jeffrey and Yang, Yinfei and Gan, Zhe},
  booktitle={European Conference on Computer Vision},
  pages={240--255},
  year={2025},
  organization={Springer}
}

@article{gu2024middleware,
  title={Middleware for llms: Tools are instrumental for language agents in complex environments},
  author={Gu, Yu and Shu, Yiheng and Yu, Hao and Liu, Xiao and Dong, Yuxiao and Tang, Jie and Srinivasa, Jayanth and Latapie, Hugo and Su, Yu},
  journal={arXiv preprint arXiv:2402.14672},
  year={2024}
}

@article{wang2406mixture,
  title={Mixture-of-agents enhances large language model capabilities},
  author={Wang, Junlin and Wang, Jue and Athiwaratkun, Ben and Zhang, Ce and Zou, James},
  year={2024},
  journal={arXiv preprint arXiv:2406.04692}
}

@article{wu2023autogen,
  title={Autogen: Enabling next-gen llm applications via multi-agent conversation framework},
  author={Wu, Qingyun and Bansal, Gagan and Zhang, Jieyu and Wu, Yiran and Zhang, Shaokun and Zhu, Erkang and Li, Beibin and Jiang, Li and Zhang, Xiaoyun and Wang, Chi},
  journal={arXiv preprint arXiv:2308.08155},
  year={2023}
}

@article{zhou2309agents,
  title={Agents: An open-source framework for autonomous language agents},
  author={Zhou, Wangchunshu and Jiang, Yuchen Eleanor and Li, Long and Wu, Jialong and Wang, Tiannan and Qiu, Shi and Zhang, Jintian and Chen, Jing and Wu, Ruipu and Wang, Shuai and others},
  journal={arXiv preprint arXiv:2309.03409},
  year={2023},
}

@article{achiam2023gpt,
  title={Gpt-4 technical report},
  author={Achiam, Josh and Adler, Steven and Agarwal, Sandhini and Ahmad, Lama and Akkaya, Ilge and Aleman, Florencia Leoni and Almeida, Diogo and Altenschmidt, Janko and Altman, Sam and Anadkat, Shyamal and others},
  journal={arXiv preprint arXiv:2303.08774},
  year={2023}
}

@inproceedings{jansen2005software,
  title={Software architecture as a set of architectural design decisions},
  author={Jansen, Anton and Bosch, Jan},
  booktitle={5th Working IEEE/IFIP Conference on Software Architecture (WICSA'05)},
  pages={109--120},
  year={2005},
  organization={IEEE}
}

@inproceedings{lago2024threats,
  title={Threats to validity in software engineering--hypocritical paper section or essential analysis?},
  author={Lago, Patricia and Runeson, Per and Song, Qunying and Verdecchia, Roberto},
  booktitle={Proceedings of the 18th ACM/IEEE International Symposium on Empirical Software Engineering and Measurement},
  pages={314--324},
  year={2024}
}

@inproceedings{park2023generative,
  title={Generative agents: Interactive simulacra of human behavior},
  author={Park, Joon Sung and O'Brien, Joseph and Cai, Carrie Jun and Morris, Meredith Ringel and Liang, Percy and Bernstein, Michael S},
  booktitle={Proceedings of the 36th Annual ACM Symposium on User Interface Software and Technology},
  pages={1--22},
  year={2023}
}

@article{wang2022self,
  title={Self-consistency improves chain of thought reasoning in language models},
  author={Wang, Xuezhi and Wei, Jason and Schuurmans, Dale and Le, Quoc and Chi, Ed and Narang, Sharan and Chowdhery, Aakanksha and Zhou, Denny},
  journal={arXiv preprint arXiv:2203.11171},
  year={2022}
}

@article{wang2023rcagent,
  title={RCAgent: Cloud Root Cause Analysis by Autonomous Agents with Tool-Augmented Large Language Models},
  author={Wang, Zefan and Liu, Zichuan and Zhang, Yingying and Zhong, Aoxiao and Fan, Lunting and Wu, Lingfei and Wen, Qingsong},
  journal={arXiv preprint arXiv:2310.16340},
  year={2023}
}

@article{ye2023proagent,
  title={Proagent: From robotic process automation to agentic process automation},
  author={Ye, Yining and Cong, Xin and Tian, Shizuo and Cao, Jiannan and Wang, Hao and Qin, Yujia and Lu, Yaxi and Yu, Heyang and Wang, Huadong and Lin, Yankai and others},
  journal={arXiv preprint arXiv:2311.10751},
  year={2023}
}

@article{qiao2023making,
  title={Making language models better tool learners with execution feedback},
  author={Qiao, Shuofei and Gui, Honghao and Lv, Chengfei and Jia, Qianghuai and Chen, Huajun and Zhang, Ningyu},
  journal={arXiv preprint arXiv:2305.13068},
  year={2023}
}

@article{qu2024tool,
  title={Tool Learning with Large Language Models: A Survey},
  author={Qu, Changle and Dai, Sunhao and Wei, Xiaochi and Cai, Hengyi and Wang, Shuaiqiang and Yin, Dawei and Xu, Jun and Wen, Ji-Rong},
  journal={arXiv preprint arXiv:2405.17935},
  year={2024}
}

@misc{GoogleAI2024,
  title = {Empowering users in digital privacy management through interactive llm-based agents},
  author = {Google Research},
  year = {2024},
  url = {},
  note = {arXiv preprint arXiv:2410.11906}
}

@article{xia2023towards,
  title={Towards autonomous system: flexible modular production system enhanced with large language model agents},
  author={Xia, Y. and Shenoy, M. and Jazdi, N. and Weyrich, M.},
  journal={arXiv preprint arXiv:2304.14721},
  year={2023}
}

@article{zhao2023seehow,
  title={SeeHow: Workflow extraction from programming screencasts through action-aware video analytics},
  author={Zhao, D. and Xing, Z. and Xia, X. and Ye, D. and Xu, X. and Zhu, L.},
  journal={arXiv preprint arXiv:2304.14042},
  year={2023}
}

@article{peng2023self,
  title={Self-driven grounding: Large language model agents with automatical language-aligned skill learning},
  author={Peng, Shaohui and Hu, Xing and Yi, Qi and Zhang, Rui and Guo, Jiaming and Huang, Di and Tian, Zikang and Chen, Ruizhi and Du, Zidong and Guo, Qi and others},
  journal={arXiv preprint arXiv:2309.01352},
  year={2023}
}

@article{deng2023mind2web,
  title={Mind2web: Towards a generalist agent for the web},
  author={Deng, Xiang and Gu, Yu and Zheng, Boyuan and Chen, Shijie and Stevens, Sam and Wang, Boshi and Sun, Huan and Su, Yu},
  journal={Advances in Neural Information Processing Systems},
  volume={36},
  pages={28091--28114},
  year={2023}
}

@article{shinn2024reflexion,
  title={Reflexion: Language agents with verbal reinforcement learning},
  author={Shinn, Noah and Cassano, Federico and Gopinath, Ashwin and Narasimhan, Karthik and Yao, Shunyu},
  journal={Advances in Neural Information Processing Systems},
  volume={36},
  year={2024}
}

@article{mabrouk2023ensemble,
  title={Ensemble Federated Learning: An approach for collaborative pneumonia diagnosis},
  author={Mabrouk, Alhassan and Redondo, Rebeca P D{\'\i}az and Abd Elaziz, Mohamed and Kayed, Mohammed},
  journal={Applied Soft Computing},
  volume={144},
  pages={110500},
  year={2023},
  publisher={Elsevier}
}

@article{kang2023grounding,
  title={Grounding foundation models through federated transfer learning: A general framework},
  author={Kang, Yan and Fan, Tao and Gu, Hanlin and Fan, Lixin and Yang, Qiang},
  journal={arXiv preprint arXiv:2311.17431},
  year={2023}
}

@article{yi2024fedmoe,
  title={FedMoE: Data-Level Personalization with Mixture of Experts for Model-Heterogeneous Personalized Federated Learning},
  author={Yi, Liping and Yu, Han and Ren, Chao and Zhang, Heng and Wang, Gang and Liu, Xiaoguang and Li, Xiaoxiao},
  journal={arXiv preprint arXiv:2402.01350},
  year={2024}
}

@article{schulhoff2024prompt,
  title={The Prompt Report: A Systematic Survey of Prompting Techniques},
  author={Schulhoff, Sander and Ilie, Michael and Balepur, Nishant and Kahadze, Konstantine and Liu, Amanda and Si, Chenglei and Li, Yinheng and Gupta, Aayush and Han, HyoJung and Schulhoff, Sevien and others},
  journal={arXiv preprint arXiv:2406.06608},
  year={2024}
}

@misc{chanharms,
  title={Harms from increasingly agentic algorithmic systems (2023)},
  author={Chan, A and Salganik, R and Markelius, A and Pang, C and Rajkumar, N and Krasheninnikov, D and Langosco, L and He, Z and Duan, Y and Carroll, M and others}
}

@article{wang2024survey,
  title={A survey on large language model based autonomous agents},
  author={Wang, Lei and Ma, Chen and Feng, Xueyang and Zhang, Zeyu and Yang, Hao and Zhang, Jingsen and Chen, Zhiyuan and Tang, Jiakai and Chen, Xu and Lin, Yankai and others},
  journal={Frontiers of Computer Science},
  volume={18},
  number={6},
  pages={186345},
  year={2024},
  publisher={Springer}
}

@article{zhao2023brain,
  title={A brain-inspired theory of mind spiking neural network improves multi-agent cooperation and competition},
  author={Zhao, Zhuoya and Zhao, Feifei and Zhao, Yuxuan and Zeng, Yi and Sun, Yinqian},
  journal={Patterns},
  volume={4},
  number={8},
  year={2023},
  publisher={Elsevier}
}

@article{mei2024aios,
  title={Aios: Llm agent operating system},
  author={Mei, Kai and Li, Zelong and Xu, Shuyuan and Ye, Ruosong and Ge, Yingqiang and Zhang, Yongfeng},
  journal={arXiv preprint arXiv:2403.16971},
  year={2024}
}

@article{touvron2023llama,
  title={Llama: Open and efficient foundation language models},
  author={Touvron, Hugo and Lavril, Thibaut and Izacard, Gautier and Martinet, Xavier and Lachaux, Marie-Anne and Lacroix, Timoth{\'e}e and Rozi{\`e}re, Baptiste and Goyal, Naman and Hambro, Eric and Azhar, Faisal and others},
  journal={arXiv preprint arXiv:2302.13971},
  year={2023}
}

@article{schick2023toolformer,
  title={Toolformer: Language models can teach themselves to use tools},
  author={Schick, Timo and Dwivedi-Yu, Jane and Dess{\`\i}, Roberto and Raileanu, Roberta and Lomeli, Maria and Hambro, Eric and Zettlemoyer, Luke and Cancedda, Nicola and Scialom, Thomas},
  journal={Advances in Neural Information Processing Systems},
  volume={36},
  pages={68539--68551},
  year={2023}
}

@article{masterman2024landscape,
  title={The Landscape of Emerging AI Agent Architectures for Reasoning, Planning, and Tool Calling: A Survey},
  author={Masterman, Tula and Besen, Sandi and Sawtell, Mason and Chao, Alex},
  journal={arXiv preprint arXiv:2404.11584},
  year={2024}
}

@article{xu2025mantra,
  title={Mantra: Enhancing automated method-level refactoring with contextual rag and multi-agent llm collaboration},
  author={Xu, Yisen and Lin, Feng and Yang, Jinqiu and Tsantalis, Nikolaos and others},
  journal={arXiv preprint arXiv:2503.14340},
  year={2025}
}

@inproceedings{gandhi2024budgetmlagent,
  title={Budgetmlagent: A cost-effective llm multi-agent system for automating machine learning tasks},
  author={Gandhi, Shubham and Patwardhan, Manasi and Vig, Lovekesh and Shroff, Gautam},
  booktitle={Proceedings of the 4th International Conference on AI-ML Systems},
  pages={1--9},
  year={2024}
}

@article{budler2023review,
  title={Review of artificial intelligence-based question-answering systems in healthcare},
  author={Budler, Leona Cilar and Gosak, Lucija and Stiglic, Gregor},
  journal={Wiley Interdisciplinary Reviews: Data Mining and Knowledge Discovery},
  volume={13},
  number={2},
  pages={e1487},
  year={2023},
  publisher={Wiley Online Library}
}

@article{longo2024explainable,
  title={Explainable Artificial Intelligence (XAI) 2.0: A manifesto of open challenges and interdisciplinary research directions},
  author={Longo, Luca and Brcic, Mario and Cabitza, Federico and Choi, Jaesik and Confalonieri, Roberto and Del Ser, Javier and Guidotti, Riccardo and Hayashi, Yoichi and Herrera, Francisco and Holzinger, Andreas and others},
  journal={Information Fusion},
  volume={106},
  pages={102301},
  year={2024},
  publisher={Elsevier}
}

@article{glass1995contemporary,
  title={Contemporary application-domain taxonomies},
  author={Glass, Robert L and Vessey, Iris},
  journal={IEEE software},
  volume={12},
  number={4},
  pages={63--76},
  year={1995},
  publisher={IEEE}
}

@article{wheaton1968development,
  title={Development of a Taxonomy of Human Performance: A Review of Classificatory Systems Relating},
  author={Wheaton, George R},
  year={1968},
  publisher={Citeseer}
}

@article{prieto1991implementing,
  title={Implementing faceted classification for software reuse},
  author={Prieto-Diaz, Ruben},
  journal={Communications of the ACM},
  volume={34},
  number={5},
  pages={88--97},
  year={1991},
  publisher={ACM New York, NY, USA}
}

@article{kwasnik1999role,
  title={The role of classification in knowledge representation and discovery},
  author={Kwasnik, Barbara H},
  year={1999},
  publisher={Graduate School of Library and Information Science. University of Illinois~…}
}

@article{kamoi2024can,
  title={When can llms actually correct their own mistakes? a critical survey of self-correction of llms},
  author={Kamoi, Ryo and Zhang, Yusen and Zhang, Nan and Han, Jiawei and Zhang, Rui},
  journal={Transactions of the Association for Computational Linguistics},
  volume={12},
  pages={1417--1440},
  year={2024},
  publisher={MIT Press 255 Main Street, 9th Floor, Cambridge, Massachusetts 02142, USA~…}
}

@article{hao2024training,
  title={Training large language models to reason in a continuous latent space},
  author={Hao, Shibo and Sukhbaatar, Sainbayar and Su, DiJia and Li, Xian and Hu, Zhiting and Weston, Jason and Tian, Yuandong},
  journal={arXiv preprint arXiv:2412.06769},
  year={2024}
}

@article{valmeekam2409llms,
  title={Llms still can’t plan; can lrms? a preliminary evaluation of openai’s o1 on planbench},
  author={Valmeekam, Karthik and Stechly, Kaya and Kambhampati, Subbarao},
  journal={URL https://arxiv. org/abs/2409.13373},
  year={2024}
}

@article{zhang2024large,
  title={Large language model-brained gui agents: A survey},
  author={Zhang, Chaoyun and He, Shilin and Qian, Jiaxu and Li, Bowen and Li, Liqun and Qin, Si and Kang, Yu and Ma, Minghua and Lin, Qingwei and Rajmohan, Saravan and others},
  journal={arXiv preprint arXiv:2411.18279},
  year={2024}
}

@article{xu2024fedmllm,
  title={FedMLLM: Federated Fine-tuning MLLM on Multimodal Heterogeneity Data},
  author={Xu, Binqian and Shu, Xiangbo and Mei, Haiyang and Xie, Guosen and Fernando, Basura and Shou, Mike Zheng and Tang, Jinhui},
  journal={arXiv preprint arXiv:2411.14717},
  year={2024}
}

@misc{taskade2024top,
  author       = {Taskade},
  title        = {Top 11 Open-Source Autonomous Agents \& Frameworks: The Future of Self-Running AI},
  year         = {2024},
  url          = {https://www.taskade.com/blog/top-autonomous-agents/},
  note         = {Accessed: 2024-05-28}
}

@misc{crewai2023,
  author       = {CrewAI},
  title = {CrewAI},
  year         = {2024},
  url          = {https://www.crewai.com/},
  note         = {Accessed: 2024-05-28}
}

\appendix
\section{Questionnaire and Tables}
\label{app1}
\includepdf[pages=-]{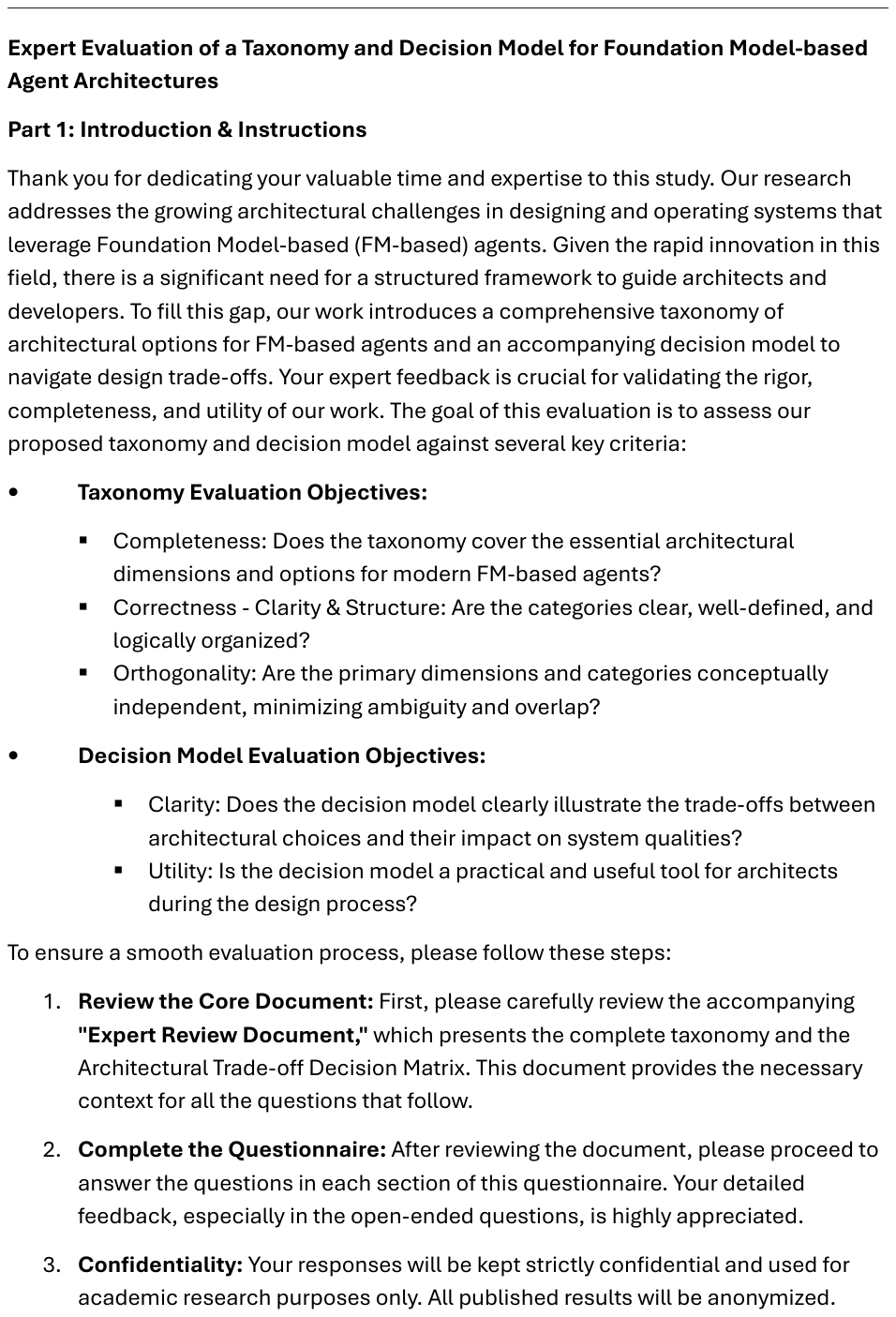}

\begin{table*}[htbp]
\centering
\caption{Decision Matrix: Impact of Design Options on Non-functional Qualities}
\label{tab:decision_matrix_final}
\resizebox{0.58\textwidth}{!}{
\begin{tabular}{@{}l|ccccccccc@{}}
\toprule
\textbf{Design Option} & \textbf{Perf.} & \textbf{Rel.} & \textbf{Rob.} & \textbf{Resil.} & \textbf{Sec.} & \textbf{Maint.} & \textbf{RAI} & \textbf{Usab.} & \textbf{Comp.} \\
\midrule
Single-modality & ++ & + & o & o & o & ++ & o & - & o \\
Multi-modality & -- & - & + & + & - & -- & + & ++ & - \\
\midrule
Interactional Context & - & o & o & o & o & - & - & + & o \\
Environmental Context & o & + & o & o & o & o & o & + & o \\
Semantic Context & o & ++ & o & o & o & o & + & ++ & o \\
Passive Suggestion & + & o & + & o & o & o & + & o & - \\
Proactive Suggestion & - & + & - & + & o & - & - & - & ++ \\
\midrule
Agent-as-a-coordinator & - & - & o & - & + & + & + & o & o \\
Agent-as-a-worker & + & + & o & o & o & ++ & o & o & o \\
\midrule
Task Completion & + & o & o & o & o & + & o & + & o \\
Communication & - & - & + & + & -- & - & ++ & ++ & ++ \\
Learning & - & -- & ++ & ++ & - & -- & -- & + & o \\
\midrule
Internal Planner & + & o & o & o & o & o & + & o & o \\
External Planner & - & ++ & ++ & o & o & + & - & o & o \\
Hybrid Approach & o & + & + & o & o & - & o & o & + \\
Internal Tool & o & o & - & o & - & - & - & o & o \\
External Tool & - & ++ & ++ & o & o & + & - & o & o \\
Single-path Plan Generator & + & - & - & - & o & ++ & + & - & o \\
Multi-path Plan Generator & - & o & ++ & ++ & o & - & o & ++ & o \\
\midrule
Long-term Memory & o & + & o & + & o & o & - & + & + \\
Short-term Memory & ++ & o & + & o & o & - & ++ & o & ++ \\
Selective Forgetting & - & + & + & o & o & + & - & ++ & o \\
Natural Language Memory & - & o & o & o & o & o & ++ & ++ & o \\
Embedding Memory & ++ & o & o & o & o & - & - & - & o \\
Databases & o & ++ & ++ & ++ & ++ & ++ & + & + & o \\
Structured Lists & ++ & + & ++ & + & o & o & ++ & + & o \\
Memory Reading & + & -- & o & o & o & o & o & + & o \\
Memory Writing & + & -- & o & o & o & o & o & o & o \\
Memory Reflection & - & ++ & + & o & o & o & - & + & o \\
Memory Sharing & - & + & - & o & o & - & - & + & o \\
\midrule
Centralized & + & - & - & - & + & + & + & o & o \\
Federated & o & + & o & o & o & - & o & o & o \\
Fully Distributed & - & ++ & ++ & ++ & - & -- & -- & o & o \\
Full Transparency & + & + & o & - & o & - & o & + & o \\
Partial Transparency & o & o & o & o & o & + & o & o & o \\
Configuration-free & ++ & o & + & o & o & o & ++ & o & ++ \\
Customizable & - & + & o & o & o & o & - & o & - \\
API & ++ & ++ & + & + & + & + & + & o & ++ \\
UI Understanding & -- & o & o & o & o & o & o & ++ & -- \\
API-based Learning & ++ & + & o & o & o & + & o & - & ++ \\
UI-based Learning & -- & -- & -- & -- & - & -- & - & ++ & ++ \\
Inline Integration & + & o & o & o & o & o & + & o & o \\
Multi-source Integration & - & + & o & o & o & o & - & o & o \\
Physical Actuation & o & o & -- & o & o & -- & o & -- & o \\
Virtual Actuation & o & + & o & o & o & + & o & o & o \\
\midrule
Workflow/Plan Generation & o & o & o & o & o & o & + & o & o \\
Intermediate Result & o & o & o & o & o & o & o & o & o \\
Final Output & -- & o & o & o & o & o & ++ & ++ & o \\
Self-reflection & - & + & + & + & o & - & + & o & o \\
Cross-reflection & - & o & o & o & - & - & + & o & o \\
Human Reflection & -- & ++ & ++ & ++ & + & o & ++ & o & o \\
Environmental Reflection & + & + & o & o & o & + & + & o & o \\
Chain of Thought (CoT) & - & + & o & o & o & + & ++ & o & o \\
Tree of Thoughts (ToT) & -- & o & o & o & o & -- & o & o & o \\
\midrule
Self-learning & o & o & o & - & o & - & - & o & o \\
In-context Learning & ++ & - & - & o & -- & ++ & + & ++ & o \\
Group Learning & o & o & o & o & o & o & + & o & o \\
\midrule
Centralized Workflow & + & o & - & o & o & o & + & o & o \\
Decentralized Workflow & - & + & ++ & ++ & ++ & - & -- & o & o \\
\midrule
Domain-specific AI & ++ & + & o & o & o & o & o & - & -- \\
General Purpose AI & - & o & + & o & o & o & o & ++ & ++ \\
Single Model & o & o & - & o & - & o & + & - & o \\
Multi-model & -- & ++ & + & + & o & -- & + & o & - \\
Sovereign AI & o/- & ++ & + & o & ++ & + & ++ & o & -- \\
Partially Sourced AI & + & o & o & o & + & o & + & o & + \\
Fully Externally Sourced AI & ++ & -- & - & -- & -- & -- & -- & o & ++ \\
Centralized Training & -- & o & o & o & + & + & - & o & o \\
Collaborative Learning & - & o & o & o & - & - & ++ & o & o/- \\
\bottomrule
\end{tabular}
}
\begin{flushleft}
\scriptsize{\textbf{Note}: `++` (strong positive), `+` (positive), `o` (neutral/unspecified), `-` (negative), `--` (strong negative). \\
\textbf{Perf.}: Performance; \textbf{Rel.}: Reliability; \textbf{Rob.}: Robustness; \textbf{Resil.}: Resilience; \textbf{Sec.}: Security; \textbf{Maint.}: Maintainability; \textbf{RAI}: Responsible \& Trustworthy AI; \textbf{Usab.}: Usability; \textbf{Comp.}: Compatibility.}
\end{flushleft}
\end{table*}

\begin{table*}[htbp]
\centering
\caption{Mapping Table -- Taxonomy Branches to Literature References}
\label{tab:mapping_table}
\resizebox{0.7\textwidth}{!}{
\begin{tabular}{@{}l|c@{}}
\toprule
\textbf{Taxonomy Branches} & \textbf{References} \\
\midrule
Input Modality $>$ Single-modality & [32, 33] \\
Input Modality $>$ Multi-modality & [34, 35, 36] \\

Context $>$ Context Types $>$ Interactional Context & [37] \\
Context $>$ Context Types $>$ Environmental Context & [38] \\
Context $>$ Context Types $>$ Semantic Context & [39] \\
Context $>$ Goal Seeking $>$ Passive Suggestion & [35] \\
Context $>$ Goal Seeking $>$ Proactive Suggestion & [40] \\

Role-playing $>$ Agent-as-a-coordinator & [1, 5, 10, 41] \\
Role-playing $>$ Agent-as-a-worker & [34, 42, 43, 44, 45, 46] \\

Goal Type $>$ Task Completion & [4, 47, 48] \\
Goal Type $>$ Communication & [48] \\
Goal Type $>$ Learning & [4] \\

Planning $>$ Planning Engine $>$ Internal Planner & [51] \\
Planning $>$ Planning Engine $>$ External Planner & [52, 53] \\
Planning $>$ Planning Engine $>$ Hybrid Approach & [53] \\
Planning $>$ Tool Selection $>$ Internal & [4, 54] \\
Planning $>$ Tool Selection $>$ External & [54] \\
Planning $>$ Plan Generation $>$ Single-path Plan Generator & [56, 57] \\
Planning $>$ Plan Generation $>$ Multi-path Plan Generator & [51, 58, 59, 60] \\

Memory $>$ Memory Types $>$ Long-term Memory & [57, 61] \\
Memory $>$ Memory Types $>$ Short-term Memory & [62] \\
Memory $>$ Memory Types $>$ Selective Forgetting & [62] \\
Memory $>$ Memory Format $>$ Natural Language Memory & [4, 63] \\
Memory $>$ Memory Format $>$ Embedding Memory & [64] \\
Memory $>$ Memory Format $>$ Databases & [48] \\
Memory $>$ Memory Format $>$ Structured Lists & [39, 62] \\
Memory $>$ Memory Operation $>$ Memory Reading & [42, 43, 46, 62] \\
Memory $>$ Memory Operation $>$ Memory Writing & [65] \\
Memory $>$ Memory Operation $>$ Memory Reflection & [62] \\
Memory $>$ Memory Operation $>$ Memory Sharing & [12, 61, 66, 67] \\

Action $>$ Agent Coordination $>$ Centralized Coordination & [12, 16] \\
Action $>$ Agent Coordination $>$ Federated Coordination & [5, 42, 43, 46, 68] \\
Action $>$ Agent Coordination $>$ Fully Distributed Coordination & [16, 47, 62] \\

Action $>$ Agent Communication $>$ Full Transparency & [57, 58, 67, 69] \\
Action $>$ Agent Communication $>$ Partial Transparency $>$ Goal-based Sharing & [16] \\
Action $>$ Agent Communication $>$ Partial Transparency $>$ Role-based Sharing & [16] \\
Action $>$ Agent Communication $>$ Partial Transparency $>$ Sensitive Data Withholding & [69, 70] \\
Action $>$ Agent Communication $>$ Partial Transparency $>$ Context-aware Sharing & [70, 71, 72] \\

Action $>$ Tool Calling $>$ Tool Invocation $>$ Configuration-free & [73] \\
Action $>$ Tool Calling $>$ Tool Invocation $>$ Customizable & [73, 74, 75] \\
Action $>$ Tool Calling $>$ Tool Interface $>$ API & [17, 54, 73, 76, 77, 78] \\
Action $>$ Tool Calling $>$ Tool Interface $>$ UI Understanding & [74, 79, 80] \\
Action $>$ Tool Calling $>$ Tool User Type $>$ User Experience Driven & [81, 82] \\
Action $>$ Tool Calling $>$ Tool User Type $>$ Agent Experience Driven & [50] \\
Action $>$ Tool Calling $>$ Tool Learning $>$ API-based Learning & [54, 73, 77] \\
Action $>$ Tool Calling $>$ Tool Learning $>$ UI-based Learning & [35, 74] \\
Action $>$ Tool Calling $>$ Output Integration $>$ Inline Integration & [54, 73] \\
Action $>$ Tool Calling $>$ Output Integration $>$ Multi-source Integration & [76] \\

Action $>$ Actuation $>$ Physical Actuation & [38, 84, 85] \\
Action $>$ Actuation $>$ Virtual Actuation & [74, 78, 86, 87] \\

Reflection $>$ Artifacts $>$ Workflow/Plan Generation & [88] \\
Reflection $>$ Artifacts $>$ Intermediate Result & [44] \\
Reflection $>$ Artifacts $>$ Final Output & [44] \\
Reflection $>$ Provider $>$ Self-reflection & [4, 34, 41, 43, 44, 63, 89] \\
Reflection $>$ Provider $>$ Cross-reflection & [58, 63] \\
Reflection $>$ Provider $>$ Human Reflection & [52, 68, 85] \\
Reflection $>$ Provider $>$ Environmental Reflection & [4, 41, 46, 49, 78, 90] \\
Reflection $>$ Mechanisms $>$ Chain of Thought (CoT) & [59, 77, 91] \\
Reflection $>$ Mechanisms $>$ Tree of Thoughts (ToT) & [51, 60] \\

Learning Capability $>$ Self-learning & [61, 63, 92, 93] \\
Learning Capability $>$ In-context Learning & [51, 52, 57, 62, 67, 94] \\
Learning Capability $>$ Group Learning $>$ Peer-to-peer Learning & [67, 72] \\
Learning Capability $>$ Group Learning $>$ Hierarchical Learning & [58] \\

Workflow $>$ Centralized Workflow & [12] \\
Workflow $>$ Decentralized Workflow & [47] \\

Access to Underlying Models $>$ Underlying Model Types $>$ Domain-specific AI & [95] \\
Access to Underlying Models $>$ Underlying Model Types $>$ General Purpose AI & [32, 93] \\
Access to Underlying Models $>$ Model Composition $>$ Single Model & [16] \\
Access to Underlying Models $>$ Model Composition $>$ Multi-model & [38, 42, 58, 96, 97, 98, 99] \\
Access to Underlying Models $>$ Model Sourcing $>$ Sovereign AI & [32] \\
Access to Underlying Models $>$ Model Sourcing $>$ Partially Sourced AI & [95] \\
Access to Underlying Models $>$ Model Sourcing $>$ Fully Externally Sourced AI & [77] \\
Access to Underlying Models $>$ Training Method $>$ Centralized Training & [32, 78, 86] \\
Access to Underlying Models $>$ Training Method $>$ Collaborative Learning & [67, 100] \\

\bottomrule
\end{tabular}
}
\end{table*}

\end{document}